\documentclass[
  10pt,
  twocolumn,
  twoside,
]{article}

\usepackage[numbers,round,comma,sort&compress]{natbib}
\setcitestyle{square}

\usepackage[
  left=0.65in,
  right=0.65in,
  top=0.65in,
  bottom=0.65in,
]{geometry}

\usepackage{titlesec}
\titleformat{\section}{\raggedright\large\bfseries}{\thesection}{1em}{}
\titleformat{\subsection}{\raggedright\large}{\thesubsection}{1em}{}
\usepackage{ragged2e}

\usepackage[font=small,labelfont=bf]{caption}

\usepackage{fancyhdr}
\usepackage{amssymb}
\usepackage{amsmath}

\usepackage{url}
\PassOptionsToPackage{hyphens}{url}\usepackage{hyperref}
\makeatletter
\g@addto@macro{\UrlBreaks}{\UrlOrds}
\makeatother

\usepackage{colortbl}

\definecolor{goodblue}{RGB}{0, 91, 187}
\usepackage{hyperref}
\hypersetup{
  colorlinks=true,
  allcolors=goodblue,
  urlcolor=goodblue,
  citecolor=goodblue,
  pdfborder={0 0 0},
  breaklinks=true,
}

\usepackage{stmaryrd}

\usepackage[normalem]{ulem}

\usepackage{textcomp}

\usepackage[export]{adjustbox}

\makeatletter

\def\CT@@do@color{%
  \global\let\CT@do@color\relax
  \@tempdima\wd\z@
  \advance\@tempdima\@tempdimb
  \advance\@tempdima\@tempdimc
  \advance\@tempdimb\tabcolsep
  \advance\@tempdimc\tabcolsep
  \advance\@tempdima2\tabcolsep
  \kern-\@tempdimb
  \leaders\vrule
  \hskip\@tempdima\@plus  1fill
  \kern-\@tempdimc
  \hskip-\wd\z@ \@plus -1fill }
\makeatother

\usepackage[table]{xcolor}

\definecolor{olivegreen}{rgb}{0.33333,.41961,0.18431}
\definecolor{forestgreen}{rgb}{0.13333,.5451,0.13333}

\definecolor{lightgrey}{rgb}{0.7,0.7,0.7}
\definecolor{verylightgrey}{rgb}{0.90,0.90,0.90}
\definecolor{veryverylightgrey}{rgb}{0.95,0.95,0.95}
\definecolor{grey}{rgb}{0.5,0.5,0.5}
\definecolor{darkgrey}{rgb}{0.3,0.3,0.3}
\definecolor{verydarkgrey}{rgb}{0.15,0.15,0.15}

\definecolor{headerblue}{HTML}{33367E}
\definecolor{unitednationsblue}{HTML}{4D88FF}

\definecolor{charcoal}{HTML}{36454F}
\definecolor{cinerous}{HTML}{98817B}
\definecolor{feldgrau}{HTML}{4D5D53}
\definecolor{glaucous}{HTML}{6082B6}
\definecolor{arsenic}{HTML}{3B444B}
\definecolor{xanadu}{HTML}{738678}

\definecolor{firebrick}{HTML}{B22222}
\definecolor{orangered}{HTML}{FF4500}
\definecolor{tomato}{HTML}{FF6347}

\definecolor{purpletaupe}{HTML}{3B444B}

\definecolor{headerorange}{RGB}{255,116,0}
\definecolor{headergray}{RGB}{230,230,230}

\definecolor{headerpop}{RGB}{230,230,230}

\definecolor{magmalight}{RGB}{252,251,195}
\definecolor{magmalightalt}{RGB}{250,240,184}
\definecolor{magmamedium}{RGB}{245,200,146}
\definecolor{magmadark}{RGB}{224,106,98}

\definecolor{icelight}{RGB}{223,242,244}
\definecolor{icelightalt}{RGB}{189,222,226}
\definecolor{icemedium}{RGB}{132,184,204}
\definecolor{icedark}{RGB}{103,153,191}

\definecolor{traitrowcolor}{RGB}{223,242,244}
\definecolor{traitrowcoloralt}{RGB}{189,222,226}

\definecolor{characterrowcolor}{RGB}{252,251,195}
\definecolor{characterrowcoloralt}{RGB}{250,240,184}

\definecolor{archetyperowcolor}{RGB}{255,213,212} 
\definecolor{archetyperowcoloralt}{RGB}{255,182,179} 

\definecolor{datasetrowcolor}{RGB}{232,244,234}
\definecolor{datasetrowcoloralt}{RGB}{210,231,214}

\definecolor{tableyellow}{HTML}{FFF9C4}
\definecolor{tablepurple}{HTML}{F3E5F5}

\newcommand{\done}[1]{}

\newenvironment{textblock}{\renewcommand{\item}{}\ignorespaces}{}

\usepackage{titlesec}
\titleformat*{\paragraph}{\bfseries}

\usepackage{tikz}
\usetikzlibrary{shapes}

\usepackage{changepage}

\newcommand\Loadedframemethod{default}
\usepackage[framemethod=\Loadedframemethod]{mdframed}

\mdfsetup{skipabove=\topskip,skipbelow=\topskip}

\tikzstyle{loglinetitle} =[draw=icedark, fill=icemedium!50, text=black]

\newenvironment{loglinebox}[1][]{

  \ifstrempty{#1}%
  {\mdfsetup{%
    frametitle={%
       \tikz[baseline=(current bounding box.east),outer sep=0pt]
        \node[loglinetitle, anchor=east,rectangle]
        {\strut~~#1:~~\strut};}}
  }%
  {\mdfsetup{%
     frametitle={%
       \tikz[baseline=(current bounding box.east),outer sep=0pt]
        \node[loglinetitle,anchor=east,rectangle]
        {\strut~~#1:~~\strut};}}%
   }%
   \mdfsetup{innertopmargin=5pt,linecolor=icedark,%
             linewidth=0.5pt,topline=true,
             frametitleaboveskip=\dimexpr-\ht\strutbox\relax,}
   \begin{mdframed}[backgroundcolor=icelight,nobreak=true]\relax%
     \raggedright
}{\end{mdframed}}

\tikzstyle{abstracttitle} =[draw=magmadark!75, fill=magmamedium!75, text=black]

\newenvironment{abstractbox}[1][]{

  \ifstrempty{#1}%
  {\mdfsetup{%
    frametitle={%
       \tikz[baseline=(current bounding box.east),outer sep=0pt]
        \node[abstracttitle, anchor=east,rectangle]
        {\strut~~#1:~~\strut};}}
  }%
  {\mdfsetup{%
     frametitle={%
       \tikz[baseline=(current bounding box.east),outer sep=0pt]
        \node[abstracttitle,anchor=east,rectangle]
        {\strut~~#1:~~\strut};}}%
   }%
   \mdfsetup{innertopmargin=5pt,linecolor=magmadark,%
             linewidth=0.5pt,topline=true,
             frametitleaboveskip=\dimexpr-\ht\strutbox\relax,}
   \begin{mdframed}[backgroundcolor=magmalight,nobreak=true]\relax%
     \raggedright
}{\end{mdframed}}

\tikzstyle{infotitle} =[draw=darkgrey, fill=lightgrey!50, text=black]

\usepackage{enumitem}

\tikzstyle{changelogtitle} =[draw=darkgrey, fill=lightgrey!50, text=black]

\usepackage{graphicx}
\usepackage{epsfig}
\usepackage{subcaption}

\usepackage{booktabs}

\usepackage{verbatim}

\usepackage{multirow}

\usepackage{enumerate}
\usepackage{enumitem}

\usepackage{amssymb}
\usepackage{amsmath}

\usepackage{ifthen}

\usepackage{longtable}

\usepackage{mathtools}

\newboolean{twocolswitch}

\usepackage{stackengine}

\usepackage{array}

\usepackage[export]{adjustbox}

\newcommand{\PreserveBackslash}[1]{\let\temp=\\#1\let\\=\temp}

\newcolumntype{P}[1]{>{\centering\arraybackslash}p{#1}}
\newcolumntype{M}[1]{>{\centering\arraybackslash}m{#1}}

\makeatletter
\newcommand*{\centerfloat}{%
  \parindent \z@
  \leftskip \z@ \@plus 1fil \@minus \textwidth
  \rightskip\leftskip
  \parfillskip \z@skip}
\makeatother

\newcommand{\sindex}[1]{}
\newcommand{\nindex}[1]{}

\newcommand{\www}[1]{\url{#1}}

\usepackage{lettrine}

\newcommand{\lmat}{\left[
    \begin{array}
    }
    \newcommand{\rmat}{\end{array}
  \right]
}

\newcommand{\usvsthem}{\textit{us-vs.-them}}
\newcommand{\Inpos}{\mathrm{In}_{\mathrm{pos}}}
\newcommand{\Outneg}{\mathrm{Out}_{\mathrm{neg}}}

\setboolean{twocolswitch}{true}

\usepackage{natbib}
\setcitestyle{square}
\raggedright

\usepackage{authblk}

\begin{document}

\title{\protect
Localized, unstable, and entangled: \\
Exploring `Us~vs.~Them' bias in Large Language Models


}


\renewcommand*{\Authsep}{, }
\renewcommand*{\Authand}{, }
\renewcommand*{\Authands}{, }
\renewcommand*{\Affilfont}{\normalsize\normalfont}
\renewcommand*{\Authfont}{\bfseries}
\setlength{\affilsep}{2em}

\author[1,4\thanks{tabia.prama@uvm.edu}]{Tabia Tanzin Prama}
\author[1,2]{Julia~Witte~Zimmerman}
\author[1,3]{Christopher~M.~Danforth}
\author[1,4,5,6\thanks{peter.dodds@uvm.edu}]{Peter~Sheridan~Dodds}

\affil[1]{
  Computational Story Lab,
  Vermont Advanced Computing Center,
  Vermont Complex Systems Institute,
  MassMutual Center of Excellence for Complex Systems and Data Science,
  University of Vermont,
  Burlington,
  VT 05405,
  US
}

\affil[2]{
  Computational Ethics Lab,
  University of Vermont,
  Burlington,
  VT 05405,
  US
}

\affil[3]{
  Department of Mathematics \& Statistics,
  University of Vermont,
  Burlington,
  VT 05405,
  US
  }

\affil[4]{
  Department of Computer Science,
  University of Vermont,
  Burlington,
  VT 05405,
  US
}

\affil[5]{
  Santa Fe Institute,
  1399 Hyde Park Rd,
  Santa Fe,
  NM 87501,
  US
}

\affil[6]{
  Complexity Science Hub,
  Metternichgasse 8,
  1030 Vienna,
  Austria
}

\date{\today}

\maketitle


\mbox{}

\bigskip
\bigskip
\bigskip

\hspace*{50pt}
\begin{minipage}{300pt}

  \begin{tabular}{>{\raggedright\arraybackslash}p{180pt}p{30pt}p{180pt}}

    \parbox{180pt}{    
      \begin{loglinebox}[Logline]
        \raggedright
        We identify sparse but unstable edge-level circuits underlying \textit{us-vs.-them} bias in LLMs and show that these pathways are functionally entangled with general language capabilities.

        \smallskip
      \end{loglinebox}
    }

    &
    
    &
    
    \parbox{180pt}{    
      \includegraphics[width=180pt,valign=m,frame]{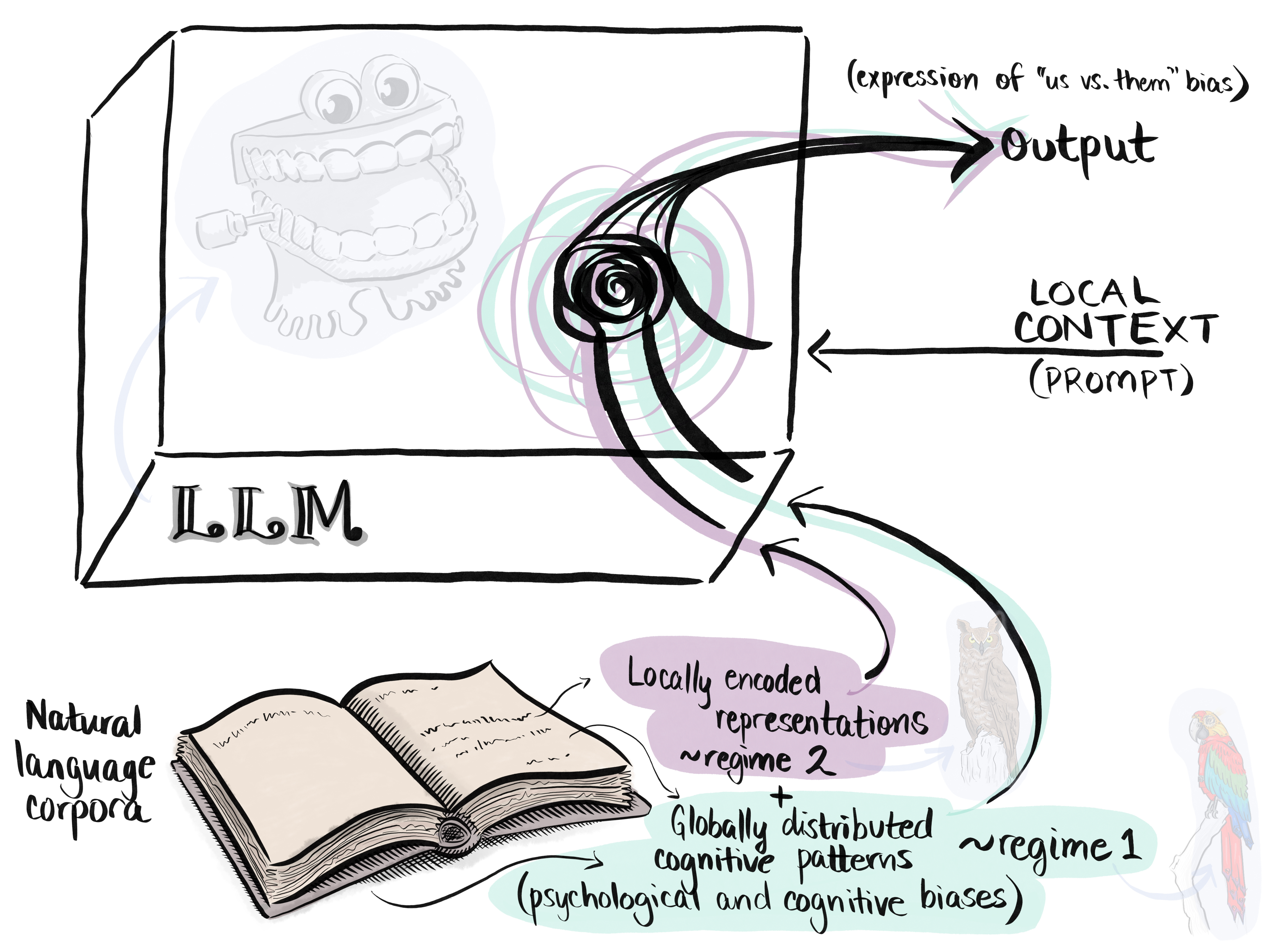}
    }
    
  \end{tabular}
  
\end{minipage}

\clearpage

\newgeometry{
  left=2in,
  right=2in,
  top=1in,
  bottom=1in,
  }

\onecolumn

\renewcommand{\baselinestretch}{1.25}
\selectfont


\begin{abstractbox}[Abstract]
  \raggedright
  \begin{textblock}
\item
  Large language models (LLMs) exhibit \textit{us-vs.-them} bias:
\item
  A behavioral asymmetry in which prompts framed around an ingroup
  (``we''/``us'')
\item
  receive systematically more positive continuations
\item
  than matched prompts framed around an outgroup
  (``they''/``them'').
\item
  Using Edge Attribution Patching (EAP),
\item
  we localize this behavior to directed circuits in
  GPT-2~Small,
  GPT-2~Large,
  Llama-2~7B,
  and
  Llama-3~8B.
\item
  Across models,
\item
  high-attribution edges are sparse
\item
  and concentrated primarily in middle-to-late layers,
\item
  while ingroup- and outgroup-related pathways are only partially overlapping
\item
  and reorganize under fine-tuning.
\item
  Behavioral comparisons and 30-token generations show
\item
  that the measured asymmetry extends beyond next-token evaluation.
\item
  Targeted patching of EAP-ranked edges reduces the bias score
\item
  more than equal-cardinality random-edge controls,
\item
  but also substantially degrades performance on
  CoLA,
  HellaSwag,
  and
  WikiText-2,
\item
  suggesting that bias-relevant pathways are functionally entangled
  with broader language capabilities.
\item
  Overall,
\item
  our findings characterize \textit{us-vs.-them} bias as a
  sparse,
  context-dependent,
  and functionally entangled computational phenomenon
\item
  whose internal organization varies across the models examined.
\end{textblock}

  \smallskip
\end{abstractbox}



\renewcommand{\baselinestretch}{1}
\selectfont

\twocolumn

\restoregeometry

\clearpage

\tableofcontents

\clearpage


\section{Introduction}

\begin{textblock}
\item
  LLMs are increasingly deployed across high-stakes domains such as
  healthcare~\cite{Nori2023CapabilitiesOG},
  education~\cite{Chu2025LLMAF},
  governance~\cite{Jeri2025TowardsAS},
  and
  content moderation~\cite{P2025ACR},
\item
  where their outputs can directly influence human decisions and social outcomes~\cite{Birhane2023ScienceIT}.
\item
  While these systems demonstrate remarkable capabilities in
  reasoning~\cite{Prama2025BanglaMATHA},
  language understanding,
  and
  generation~\cite{An2024FunAudioLLMVU},
\item
  a growing body of work has raised concerns about their tendency
  to reproduce and amplify social biases present in their training data~\cite{Prama2025UsvsThemBI,Prama2025EvaluatingCA}.
\item
  These biases span
  gender~\cite{Dong2024DisclosureAM},
  race~\cite{Iso2025EvaluatingBI},
  religion~\cite{Tom2025ReligionAR},
  and
  other demographic attributes,
\item
  raising concerns about
  stereotyping,
  inequitable treatment,
  and
  downstream harm~\cite{Gallegos2023BiasAF}.
\end{textblock}

\begin{textblock}
\item
  At the core of many social biases lies the distinction between
  ingroups (``us'')
  and
  outgroups (``them'')~\cite{Tajfel2004TheSI}.
\item
  We use \usvsthem{} bias to describe a systematic asymmetry in responses
  toward an ingroup relative to an outgroup,
\item
  expressed through
  ingroup favoritism,
  outgroup derogation,
  or both.
\item
  These processes are related but distinct:
\item
  Greater positivity toward an ingroup does not necessarily imply hostility toward an outgroup.
\item
  Research in social psychology~\cite{2004ThePO}
  and
  social identity theory~\cite{Ashforth1989SocialIT}
  shows that such asymmetries shape judgments across domains including
  race~\cite{Hippel1997TheLI},
  gender~\cite{Egan2001GenderIA},
  and
  religion~\cite{Rowatt2013ReligionPA}.
\item
  LLMs exhibit analogous output patterns,
\item
  producing more positive continuations under ingroup framing
\item
  and more negative continuations under outgroup framing~\cite{Prama2025UsvsThemBI,Hu2023GenerativeLM}.
\item
  Here, this terminology refers to measurable differences in model outputs
\item
  and does not imply that LLMs possess social identities or conscious group affiliations.
\item
  When deployed in downstream applications,
\item
  these patterns can contribute to
  unequal treatment,
  stereotyping,
  and
  misinformation~\cite{Gallegos2023BiasAF}.
\item
  However, existing evaluation approaches often focus on
  isolated demographic contrasts~\cite{Esiobu2023ROBBIERB}
  or
  aggregate sentiment measures~\cite{Prama2025EvaluatingCA},
\item
  providing limited insight into the broader ingroup--outgroup structure underlying such behavior.
\end{textblock}

\begin{textblock}
\item
  Recent advances in mechanistic interpretability have begun to open the black box of neural networks~\cite{Bereska2024MechanisticIF},
\item
  enabling researchers to identify internal components such as attention heads and MLP layers
  that contribute to specific model behaviors~\cite{Nanda2023ProgressMF}.
\item
  However, most of this work has focused on narrow linguistic phenomena or individual features~\cite{Wu2025LargeLM},
\item
  with limited attention to the mechanisms driving broader social biases across contexts and identities.
\item
  Moreover, emerging evidence suggests that behavioral alignment at the output level
  is not a reliable gauge of internal representations~\cite{Chen2025ReasoningMD,Zimmerman2024TokensTO},
\item
  motivating mechanistic analysis that looks beneath model outputs.
\item
  We therefore ask:
\item
  Can \usvsthem{} bias be localized to identifiable substructures within an LLM?
\item
  If so, this would enable targeted, component-level interventions---
  modifying specific edges or layers rather than relying on
  prompting~\cite{Gallegos2023BiasAF,Dwivedi2023BreakingTB},
  data augmentation~\cite{Zhao2018GenderBI,Zmigrod2019CounterfactualDA},
  or
  full fine-tuning~\cite{Ouyang2022TrainingLM,Shen2023LooseLS}---
  and would shed light on whether social biases as fundamental as the ingroup--outgroup distinction
\item 
  are deeply embedded in the computational structure of learned language representations.
\end{textblock}

\begin{textblock}
\item
  Among the different component types available for analysis,
\item
  we focus on \textit{edges}---directed connections between computational units such as attention heads and MLP layers---
\item
  and quantify their causal contribution using Edge Attribution Patching (EAP)~\cite{Syed2023AttributionPO}.
\item
  Each edge represents the flow of information between adjacent stages of the transformer's forward pass;
\item
  its importance is estimated by measuring how much the bias signal changes
  when that edge's activation is replaced by a corrupted counterpart~\cite{Bhaskar2024FindingTC}.
\item
  This edge-centric view allows us to move beyond node-level attribution
  and construct directed bias circuits~\cite{Gu2025DiscoveringTC}:
\item
  Sparse subgraphs that are causally sufficient for the observed biased behavior~\cite{Miller2024TransformerCF}.
\item
  Our main contributions are as follows:
\end{textblock}

\noindent\textbf{RQ1. Localized encoding of \usvsthem{} bias in LLM circuits.}
\begin{textblock}
\item
  We conduct a systematic analysis across four open-source LLMs
  (GPT-2~Small, GPT-2~Large, Llama-2~7B, and Llama-3~8B)
\item
  to test whether \usvsthem{} bias is localized within a small subset of edges.
\item
  Using multiple circuit-identification approaches,
\item
  we find that biased behavior is concentrated in specific edges and layers across models
  (Figures~\ref{fig:four_models_edge_comparison} and~\ref{fig:ablation}).
\end{textblock}

\noindent\textbf{RQ2. Instability of bias circuits across corruption strategy and fine-tuning.}
\begin{textblock}
\item
  We evaluate the robustness of identified
  ingroup-positivity,
  outgroup-negativity,
  and
  combined \usvsthem{} edges
\item
  under variations in corruption strategy and fine-tuning condition.
\item
  Our analysis (Figure~\ref{fig:five_overlap} and Table~\ref{tab:change_in_metric_updated})
  shows that the specific edges associated with biased behavior
\item
  are not stable under perturbations to either the model or the input space.
\end{textblock}

\noindent\textbf{RQ3. Causal specificity and capability entanglement of bias-related edges.}
\begin{textblock}
\item
  We test whether EAP-identified edges have a disproportionate causal influence on the measured bias signal
  relative to randomly selected edges,
\item
  then evaluate the effect of these interventions on
  linguistic acceptability,
  commonsense continuation,
  and
  language-modeling performance.
\item
  This determines whether the identified pathways are selective to the measured bias behavior
\item
  or functionally entangled with broader language capabilities.
\end{textblock}

\section{Related Work}

\paragraph{Bias in LLMs.}
\begin{textblock}
\item
  Prior work shows that LLMs reproduce social biases related to
  gender,
  race,
  religion,
  nationality,
  age,
  social identity,
  political,
  and
  other demographic attributes~\cite{Abid2021PersistentAB,Ahn2021MitigatingLE,Kotek2023GenderBA,Kamruzzaman2023InvestigatingSB,Soundararajan2024InvestigatingGB,Prama2025EvaluatingCA}.
\item
  Similar biases were previously observed in word embeddings trained on large text corpora~\cite{Caliskan2016SemanticsDA,Bolukbasi2016QuantifyingAR}.
\item
  Common mitigation approaches include
  counterfactual data augmentation~\cite{Zhao2018GenderBI,Zmigrod2019CounterfactualDA},
  prompt-based debiasing~\cite{Gallegos2023BiasAF,Dwivedi2023BreakingTB},
  fine-tuning or alignment-based methods~\cite{Ouyang2022TrainingLM,Shen2023LooseLS},
  and
  representation-level interventions~\cite{Nizhnichenkov2026ARA}.
\item
  However, debiasing can also introduce trade-offs,
\item
  including reduced downstream performance
  or incomplete removal of biased internal representations~\cite{Kaneko2023TheIO}.
\end{textblock}

\paragraph{Mechanistic interpretability for LLMs.}
\begin{textblock}
\item
  Mechanistic interpretability (MI) aims to reverse-engineer trained neural networks
  by identifying the internal computations that give rise to model behavior,
\item
  analogous to analyzing compiled software binaries~\cite{olah2022mechinterp}.
\item
  In transformer language models,
\item
  MI studies units such as
  neurons,
  attention heads,
  MLP layers,
  residual streams,
  and
  directed edges
\item
  to explain how specific behaviors emerge.
\item
  Prior work has used circuit-based and causal methods to analyze
  induction heads,
  factual recall,
  indirect object identification,
  grokking,
  and
  information flow in transformers~\cite{Olah2020ZoomIA,Goh2021MultimodalNI,Geiger2021CausalAO,Meng2022LocatingAE,Wang2022InterpretabilityIT}.
\item
  More recent work has scaled these approaches through
  automated circuit discovery~\cite{Conmy2023TowardsAC},
  causal tests for algorithmic hypotheses~\cite{Wu2023InterpretabilityAS},
  circuit analyses of grokking~\cite{Nanda2023ProgressMF},
  refined accounts of induction heads~\cite{GoldowskyDill2023LocalizingMB},
  and
  methods for tracing information flow through transformer architectures~\cite{Katz2024BackwardLP}.
\item
  A growing body of work applies mechanistic interpretability to analyze and mitigate bias in LLMs~\cite{Vig2020InvestigatingGB,Chintam2023IdentifyingAA,Chandna2025DissectingBI,Gamboa2024ANI,Prakash2024InterpretingBI,Kim2025LinearRO}.
\item
  \citet{Vig2020InvestigatingGB} identified neurons and attention heads in GPT-2 contributing to gender bias,
\item
  while \citet{Chintam2023IdentifyingAA} used automated circuit discovery and ablations to reduce stereotype propagation.
\item
  \citet{Chandna2025DissectingBI} localized demographic and gender bias in GPT-2 and Llama-2
\item
  but showed it can shift across fine-tuning
  and overlap with components for downstream tasks.
\item
  \citet{Gamboa2024ANI} proposed a bias attribution score for multilingual token-level explanations,
\item
  and \citet{Prakash2024InterpretingBI} introduced activation and attribution patching
  to trace gender bias in LLaMA and Mistral.
\item
  \citet{Kim2025LinearRO} showed political ideology can be encoded as a linear direction in activation space.
\item
  These studies demonstrate that bias is present not only in model outputs
\item
  but also in internal components and information-flow pathways.
\end{textblock}

\section{Method}

\subsection{Localization of \usvsthem{} Bias}

\begin{textblock}
\item
  Our primary objective is to determine whether \usvsthem{} bias is encoded
  within specific components of large language models,
\item
  rather than being diffusely distributed across the entire network.
\item
  We use a mechanistic interpretability framework grounded in circuit analysis,
\item
  modeling transformer computation as a directed acyclic graph (DAG)
  to identify the subgraph responsible for the target behavior~\cite{Wang2019StructuredPO,Raukur2022TowardTA}.
\item
  Circuits are also explored in terms of nodes as ``features,''
\item
  typically defined as directions in the latent space~\cite{Cammarata2020ThreadC}.
\end{textblock}

\paragraph{Edge Attribution Patching.}
\label{sec:eap}

\begin{textblock}
\item
  To localize bias-related computation,
\item
  we use Edge Attribution Patching (EAP)~\cite{Syed2023AttributionPO},
\item
  a causal attribution method that estimates the importance of individual edges in the model's computational graph
  without requiring exhaustive ablation~\cite{Ghorbani2020NeuronSD}.
\item
  Given a clean input $x_{\mathrm{clean}}$ and a corrupted input $x_{\mathrm{corr}}$,
\item
  the causal importance of an edge $E$ is defined as the change in a bias metric $L$
  when the activation transmitted through that edge is replaced by its corrupted counterpart:
\end{textblock}
\begin{equation}
  \mathrm{Attr}(E)
  =
  \left|
    L\!\left(
      x_{\mathrm{clean}}
      \mid
      \mathrm{do}(E = e_{\mathrm{corr}})
    \right)
    -
    L(x_{\mathrm{clean}})
  \right|,
  \label{eq:eap_intervention}
\end{equation}
\begin{textblock}
\item
  where $e_{\mathrm{corr}}$ denotes the activation obtained from the corrupted input and injected at edge $E$.
\item
  A larger attribution value indicates that the edge contributes more strongly to the bias signal.
\item
  Following EAP,
\item
  we approximate edge importance using a first-order gradient-based score.
\item
  For each $(\mathrm{src}, \mathrm{dst})$ pair where $\mathrm{layer}(\mathrm{src}) < \mathrm{layer}(\mathrm{dst})$,
\item
  the directed edge score is defined as:
\end{textblock}
\[
  \mathrm{score}(\mathrm{src} \rightarrow \mathrm{dst})
  =
  \left|
    \nabla_{a_{\mathrm{dst}}} L
    \cdot
    \left(
      a_{\mathrm{src}}^{\mathrm{clean}}
      -
      a_{\mathrm{src}}^{\mathrm{corr}}
    \right)
  \right|.
\]
\begin{textblock}
\item
  Here, MLP activations and source attention heads are represented in $d_{\mathrm{model}}$ space,
\item
  with source heads taken after $W_O$,
\item
  while destination heads use pre-projection per-head value vectors $v^h \in \mathbb{R}^{d_{\mathrm{head}}}$.
\item
  This asymmetry compares source heads in a shared space
  while preserving distinct destination-head gradients.
\item
  The resulting score traces how clean--corrupted activation differences influence the bias metric,
\item
  enabling directed bias-circuit construction.
\end{textblock}

\paragraph{Bias Metric.}
\label{sec:bias_metric}

\begin{textblock}
\item
  To quantify \usvsthem{} bias,
\item
  we define sentiment-based metrics over the model's next-token distribution.
\item
  Unlike prior work, which relied on LLM-based annotations~\cite{Prama2025UsvsThemBI},
\item
  we use the English LabMT lexicon~\cite{Dodds2011TemporalPO},
\item
  which assigns happiness scores to 10,219 words on a 1--9 scale.
\item
  We define a positive lexicon $\mathcal{V}_{+}$ as words scoring above 6
\item
  and a negative lexicon $\mathcal{V}_{-}$ as words scoring below 4,
\item
  excluding intermediate-score words to keep positive and negative sentiment clearly separated.
\item
  For each prompt $x_i$,
\item
  let $\mathcal{T}_k(x_i)$ denote the top-$k$ predicted tokens at the completion position
\item
  and $p_i(t)$ the probability assigned to token $t$.
\item
  We compute the probability mass assigned to positive and negative lexical continuations as:
\end{textblock}
\begin{equation}
  \begin{aligned}
    P_{+}^{(i)}
    &=
    \sum_{t \in \mathcal{T}_k(x_i) \cap \mathcal{V}_{+}}
    p_i(t),
    &
    P_{-}^{(i)}
    &=
    \sum_{t \in \mathcal{T}_k(x_i) \cap \mathcal{V}_{-}}
    p_i(t).
  \end{aligned}
  \label{eq:positive_negative_mass}
\end{equation}
\begin{textblock}
\item
  We then define two complementary bias metrics~\cite{Chandna2025DissectingBI}.
\item
  The first captures relative sentiment polarity:
\end{textblock}
\begin{equation}
  L_1
  =
  \frac{1}{m}
  \sum_{i=1}^{m}
  \left(
    P_{+}^{(i)}
    -
    P_{-}^{(i)}
  \right),
  \label{eq:L1}
\end{equation}
\begin{textblock}
\item
  where $m$ is the number of prompts.
\item
  Higher values of $L_1$ indicate more positively valenced completions,
\item
  whereas lower values indicate more negatively valenced completions.
\end{textblock}

\begin{textblock}
\item
  The second metric isolates positive lexical association:
\end{textblock}
\begin{equation}
  L_2
  =
  \frac{1}{m}
  \sum_{i=1}^{m}
  P_{+}^{(i)}.
  \label{eq:L2}
\end{equation}
\begin{textblock}
\item
  Unlike $L_1$,
\item
  $L_2$ does not contrast positive against negative mass.
\item
  We use both metrics because ingroup positivity and outgroup negativity need not be symmetric:
\item
  Elevated positive associations for an ingroup need not coincide with elevated negative associations for an outgroup.
\end{textblock}

\begin{textblock}
\item
  We compare our LabMT-based $L_1$ metric with
  CrowS-Pairs~\cite{Nangia2020CrowSPairsAC}
  and
  StereoSet~\cite{Nadeem2020StereoSetMS}
  (see Appendix~\ref{sec:circuit}, Table~\ref{tab:external_validity}).
\item
  CrowS-Pairs and StereoSet produce identical model rankings (Spearman $\rho=1.0$)
\item
  and are negatively correlated with both $\Inpos$ and $\Outneg$ $L_1$ scores ($\rho=-1.0$, $n=4$).
\item
  For outgroup prompts,
\item
  this negative relationship indicates agreement
  because lower $L_1$ values represent stronger negative sentiment,
\item
  whereas the ingroup result suggests that ingroup positivity captures a distinct dimension of bias.
\item
  StereoSet LM scores of 96.11--97.72\% confirm that the results are not driven by incoherent completions.
\end{textblock}

\paragraph{Dataset and Corrupted Inputs.}
\label{sec:dataset_corruption}

\begin{textblock}
\item
  EAP requires paired clean and corrupted inputs
  that preserve token-level structure while perturbing the causal feature of interest.
\item
  We use Symmetric Token Replacement (STR)~\cite{Zhang2023TowardsBP}
  to maintain alignment under controlled semantic changes.
\item
  Our dataset contains 660 identity-conditioned prompts covering 165 persona traits across 16 social domains,
\item
  including
  political ideology,
  religion,
  gender identity,
  race and ethnicity,
  profession,
  socioeconomic class,
  culture,
  education,
  technology,
  values,
  law and order,
  geopolitics,
  health,
  and
  intellectual orientation.
\item
  Each identity is instantiated using four syntactic templates (Appendix~\ref{sec:dataset}).
\item
  We evaluate
  GPT-2~Small,
  GPT-2~Large,
  Llama-2~7B (\texttt{Llama-2-7b-hf}),
  and
  Llama-3~8B (\texttt{Meta-Llama-3-8B})
  on completions of these templates;
\item
  model details appear in Appendix~\ref{sec:model_details}.
\end{textblock}

\begin{textblock}
\item
  For each clean prompt,
\item
  we construct two corrupted variants.
\item
  C1 replaces the identity with an out-of-distribution placeholder such as \textit{abc},
\item
  removing identity information while preserving grammatical structure.
\item
  C2 swaps ingroup pronouns such as \textit{we} and \textit{us} with \textit{they} and \textit{them},
\item
  changing group framing while keeping the identity fixed.
\item
  Representative examples are provided in Appendix~\ref{sec:clean_and_corrupted}, Table~\ref{tab:clean_corrupt_examples}.
\end{textblock}




\subsection{Bias Circuit Localization}
\label{sec:localization_analysis}

\begin{textblock}
\item
  To assess whether \usvsthem{} bias is localized to a sparse set of edges
  or distributed broadly across the network,
\item
  we conduct two complementary analyses.
\item
  First, we evaluate circuit faithfulness
  to quantify how well the top-$K$ EAP-scored edges reproduce the full model's bias signal.
\item
  If bias is localized,
\item
  retaining only these high-attribution edges should preserve most of the bias metric.
\item
  Second, we examine the layerwise distribution of high-attribution edges
  to identify where bias-related processing is concentrated within the model architecture.
\end{textblock}

\paragraph{Circuit Faithfulness Evaluation.}
\label{sec:change_metric}

\begin{textblock}
\item
  To validate the circuits identified by EAP,
\item
  we follow prior work~\cite{Chandna2025DissectingBI}
\item
  and compute the difference between the bias metric of the full model
  and the metric obtained when only the top-$K$ edges are retained:
\end{textblock}
\begin{equation}
  \Delta
  =
  \left|
    L(\text{full model})
    -
    L(\text{top-}K\text{ edges})
  \right|.
  \label{eq:delta_metric}
\end{equation}
\begin{textblock}
\item
  Here, $L(\text{full model})$ denotes the bias metric computed over the complete network,
\item
  and $L(\text{top-}K\text{ edges})$ denotes the metric after ablating all edges except the top-$K$ edges identified by EAP.
\item
  A small value of $\Delta$ indicates that the selected edges capture most of the full model's bias signal,
\item
  providing evidence of localization.
\item
  Conversely, a large value of $\Delta$ suggests that the bias signal is distributed across a wider set of model components.
\end{textblock}


\paragraph{Circuit Instability Under Fine-Tuning.}
\label{sec:circuit_stability}

\begin{textblock}
\item
  We evaluate whether the identified bias circuits remain stable after fine-tuning under two settings.
\item
  First, we use the ingroup--outgroup neutralization (ION) procedure from prior work~\cite{Prama2025UsvsThemBI},
\item
  which combines instruction fine-tuning with Direct Preference Optimization (DPO)
  to reduce bias-related sentiment divergence.
\item
  We test whether explicit debiasing changes the internal edges responsible for bias-related completions.
\item
  Second, we fine-tune on the Shakespeare corpus,
\item
  consisting of the complete works of William Shakespeare~\cite{mohsen_shakespeare_txt}.
\item
  Because this corpus has no topical overlap with our identity-conditioned bias prompts,
\item
  it serves as a task-irrelevant fine-tuning control.
\item
  This setting tests whether the identified circuits reflect bias-specific computation
  rather than general adaptation to new language modeling data.
\item
  After each fine-tuning condition,
\item
  we recompute EAP scores
  and compare the resulting high-attribution edge sets with those from the original model.
\end{textblock}

\subsection{Causal Validation via Targeted Edge Patching}
\label{sec:inference_debiasing}

\begin{textblock}
\item
  We use targeted edge patching as a causal diagnostic
  to test whether EAP-identified edges disproportionately influence the measured \usvsthem{} bias signal.
\item
  For each clean input,
\item
  we construct a token-aligned corrupted counterpart
\item
  and replace activations along selected bias-critical edges with the corresponding corrupted activations,
\item
  while preserving clean activations along all remaining edges.
\item
  We compare this intervention with equal-cardinality random-edge controls.
\item
  This tests whether EAP-ranked edges exert greater influence on the measured bias signal than randomly selected edges
\item
  and whether their function is separable from general language competence.
\item
  We therefore evaluate changes in the bias metric together with performance on
  CoLA~\cite{warstadt2018neural},
  HellaSwag~\cite{Zellers2019HellaSwagCA},
  and
  WikiText-2~\cite{merity2016pointer}.
\item
  Selective reductions in the bias metric would indicate circuit specificity,
\item
  whereas broad performance degradation relative to random-edge controls
  would indicate functional entanglement with general language computation~\cite{Kaneko2023TheIO,Chandna2025DissectingBI}.
\end{textblock}

\begin{textblock}
\item
  We use the \texttt{HookedTransformer} class from the TransformerLens repository,\footnote{\url{https://github.com/TransformerLensOrg/TransformerLens}}
\item
  which offers a modular and transparent framework for studying the internal mechanisms of transformer models such as GPT-2 and Llama-2.
\item
  We ran all experiments on a machine with two A100 GPUs.
\end{textblock}

\section{Results and Discussion}

\paragraph{RQ1: Bias Localization.}
\label{sec:localisation}

\begin{textblock}
\item
  Appendix~\ref{sec:all_bias_circuits} shows the combined \usvsthem{} bias circuits for
  GPT-2~Small (Figure~\ref{fig:bias_circuit_gpt2_small_appendix}),
  GPT-2~Large (Figure~\ref{fig:bias_circuit_gpt2_large_appendix}),
  Llama-2~7B (Figure~\ref{fig:bias_circuit_llama2_appendix}),
  and
  Llama-3~8B (Figure~\ref{fig:bias_circuit_llama3_appendix}).
\item
  Across models,
\item
  the circuits share a broad structure---
  an early group-framing pathway,
  a middle attention-mediated routing stage,
  and
  a late evaluative region before the logits---
\item
  but become more distributed with scale.
\item
  GPT-2~Small is most compact,
\item
  with routing concentrated in layers 3--9 and readout through layers 10--11;
\item
  GPT-2~Large extends this pattern across layers 6--35,
\item
  with a larger late MLP-heavy region.
\item
  Llama-2~7B shows
  early attention--MLP interactions in layers 1--8,
  mid-layer convergence around layers 13--20,
  and
  a late hub across layers 21--31,
\item
  while Llama-3~8B has the broadest topology,
\item
  with
  an early MLP spine in layers 2--11,
  attention routing around layers 12--14,
  bottlenecks around layers 16--20,
  and
  a late module spanning layers 21--31.
\item
  These results suggest that \usvsthem{} bias is partially localized in sparse edge sets
  rather than implemented as a single fixed circuit.
\item
  The ingroup and outgroup circuits in Appendix~\ref{sec:circuit} further show
  these bias conditions are asymmetric and context-dependent
  rather than mirror images of one shared mechanism.
\end{textblock}


\begin{table*}[t!]
\centering
\begin{tabular}{|ccc|cc|cc|cc|cc|}
\hline
\rowcolor{tablepurple}
\multicolumn{3}{|c|}{}
& \multicolumn{2}{c|}{\textbf{GPT-2 Small}}
& \multicolumn{2}{c|}{\textbf{GPT-2 Large}}
& \multicolumn{2}{c|}{\textbf{Llama-2 7B}}
& \multicolumn{2}{c|}{\textbf{Llama-3 8B}} \\
\hline
\rowcolor{tablepurple}
\textbf{Bias Type} & \textbf{Metric} & \textbf{Corruption}
& \textbf{Base} & $\boldsymbol{\Delta}$
& \textbf{Base} & $\boldsymbol{\Delta}$
& \textbf{Base} & $\boldsymbol{\Delta}$
& \textbf{Base} & $\boldsymbol{\Delta}$ \\
\hline
\rowcolor{tableyellow}
$\Inpos$ & $L_1$ & $C_1$
& 0.277 & 0.168 & 0.250 & 0.214 & 0.175 & 0.142 & 0.233 & 0.150 \\
\rowcolor{tableyellow}
$\Inpos$ & $L_1$ & $C_2$
& 0.277 & 0.181 & 0.250 & 0.219 & 0.175 & 0.155 & 0.233 & 0.185 \\
\hline
$\Inpos$ & $L_2$ & $C_1$
& 0.392 & 0.192 & 0.330 & 0.277 & 0.255 & 0.201 & 0.316 & 0.219 \\
$\Inpos$ & $L_2$ & $C_2$
& 0.392 & 0.222 & 0.330 & 0.282 & 0.255 & 0.217 & 0.316 & 0.224 \\
\hline \hline
\rowcolor{tableyellow}
$\Outneg$ & $L_1$ & $C_1$
& 0.293 & 0.203 & 0.289 & 0.220 & 0.246 & 0.209 & 0.267 & 0.190 \\
\rowcolor{tableyellow}
$\Outneg$ & $L_1$ & $C_2$
& 0.293 & 0.211 & 0.289 & 0.226 & 0.246 & 0.209 & 0.267 & 0.202 \\
\hline
$\Outneg$ & $L_2$ & $C_1$
& 0.385 & 0.215 & 0.353 & 0.216 & 0.306 & 0.253 & 0.343 & 0.254 \\
$\Outneg$ & $L_2$ & $C_2$
& 0.385 & 0.240 & 0.353 & 0.222 & 0.306 & 0.252 & 0.343 & 0.268 \\
\hline
\end{tabular}
\caption[.]{
\begin{textblock}
\item
  Baseline metric values and circuit faithfulness scores across
  bias types,
  metrics,
  corruption strategies,
  and
  model architectures.
\item
  Base denotes the full-model bias signal,
\item
  while $\Delta$ measures the change between the full model and the extracted top-$K=30$ edge circuit.
\item
  Lower $\Delta$ indicates better recovery of the full-model bias signal.
\end{textblock}
}
\label{tab:change_in_metric_updated}
\end{table*}


\begin{textblock}
\item
  Table~\ref{tab:change_in_metric_updated} reports full-model baseline scores and circuit faithfulness values, $\Delta$,
  across
  bias type,
  metric,
  corruption strategy,
  and
  model architecture.
\item
  Across all models and both bias types,
\item
  $L_1$ generally yields lower $\Delta$ than $L_2$,
\item
  indicating that the relative polarity objective is more faithfully recovered by the extracted top-$K=30$ edge circuits
  than the positive-association objective alone.
\item
  Similarly, identity replacement ($C_1$) usually produces lower $\Delta$ than pronoun swapping ($C_2$),
\item
  suggesting that removing identity-specific information provides a cleaner counterfactual for circuit extraction
  than reframing group membership.
\item
  Based on these patterns,
\item
  we use $L_1/C_1$ as the canonical metric--corruption setting in subsequent experiments.
\item
  For ingroup positivity,
\item
  the lowest $\Delta$ under $L_1/C_1$ is observed in
  Llama-2~7B $(0.142)$,
  followed by
  Llama-3~8B $(0.150)$,
  GPT-2~Small $(0.168)$,
  and
  GPT-2~Large $(0.214)$.
\item
  For outgroup negativity,
\item
  Llama-3~8B achieves the lowest $\Delta$ under $L_1/C_1$ $(0.190)$,
  followed by
  GPT-2~Small $(0.203)$,
  Llama-2~7B $(0.209)$,
  and
  GPT-2~Large $(0.220)$.
\item
  These results indicate that sparse edge circuits recover a substantial portion of the full-model bias signal
  across all four architectures,
\item
  while GPT-2~Large shows the weakest faithfulness overall.
\item
  Notably, $\Delta$ values are generally higher for $\Outneg$ than for $\Inpos$,
\item
  suggesting that outgroup-negative behavior is somewhat less compactly captured by the same top-$K$ circuit size.
\item
  The top three attribution edges per bias type and model are reported in Table~\ref{tab:top_edges_bias_circuits}.
\end{textblock}

\begin{figure*}[t!]
    \centering
    \begin{subfigure}[b]{0.49\textwidth}
        \centering
        \includegraphics[width=\textwidth, trim=0 10 0 15, clip]{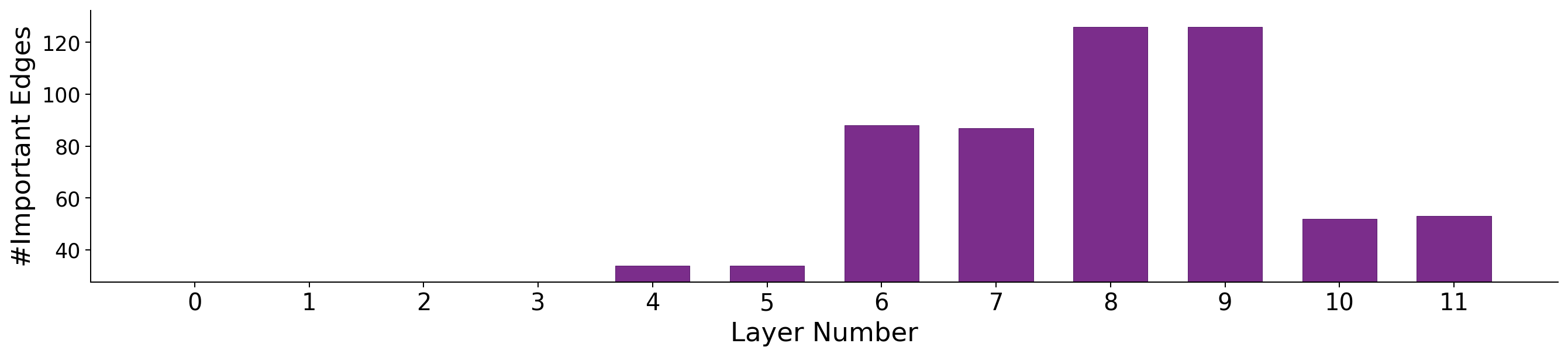}
        \caption{GPT-2 Small}
    \end{subfigure}
    \hfill
    \begin{subfigure}[b]{0.49\textwidth}
        \centering
        \includegraphics[width=\textwidth, trim=0 10 0 5, clip]{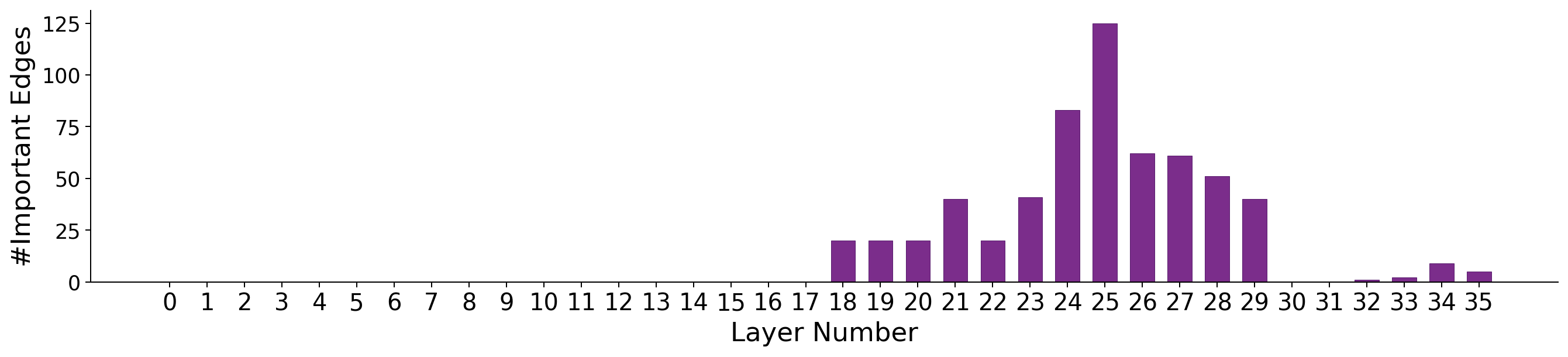}
        \caption{GPT-2 Large}
    \end{subfigure}
    \vspace{0.8em}
    \begin{subfigure}[b]{0.49\textwidth}
        \centering
        \includegraphics[width=\textwidth, trim=0 10 0 15, clip]{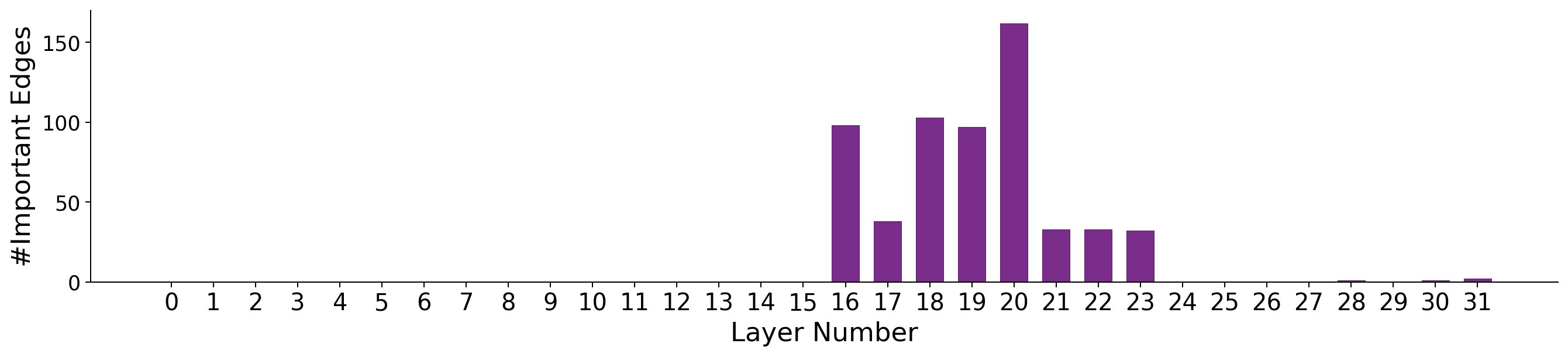}
        \caption{Llama-2 7B}
    \end{subfigure}
    \hfill
    \begin{subfigure}[b]{0.49\textwidth}
        \centering
        \includegraphics[width=\textwidth, trim=0 10 0 5, clip]{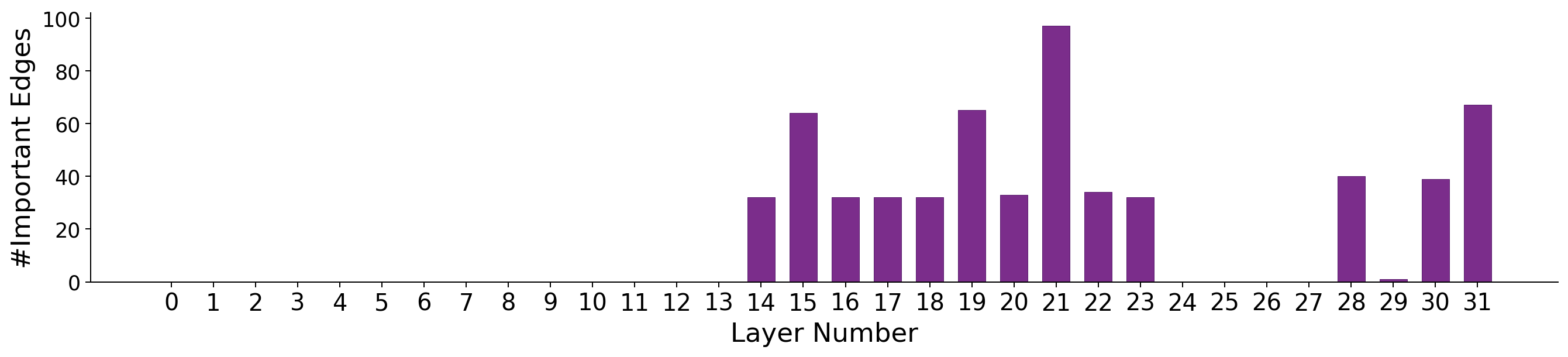}
        \caption{Llama-3 8B}
    \end{subfigure}
\caption[.]{
\begin{textblock}
\item
  Layer-wise distribution of important edges (top-$K=300$) for \usvsthem{} bias circuits across
  GPT-2~Small,
  GPT-2~Large,
  Llama-2~7B,
  and
  Llama-3~8B.
\item
  Each edge contributes to both its source and destination layer bins.
\end{textblock}
}
    \label{fig:four_models_edge_comparison}
\end{figure*}

\begin{textblock}
\item
  Figure~\ref{fig:four_models_edge_comparison} shows the layer-wise distribution of high-attribution EAP edges
  for \usvsthem{} bias across
  GPT-2~Small,
  GPT-2~Large,
  Llama-2~7B,
  and
  Llama-3~8B.
\item
  Edge counts are obtained by combining ingroup-positivity and outgroup-negativity attribution scores
  and selecting the top-ranked edges for each model.
\item
  Across architectures,
\item
  few important edges appear in early layers;
\item
  instead, bias-relevant edges concentrate in middle-to-late layers.
\item
  This suggests that the measured bias signal emerges after deeper contextual and semantic integration.
\item
  The location of these edges varies by model.
\item
  In GPT-2~Small,
\item
  important edges begin around layer 4 and peak near layers 8--9.
\item
  In GPT-2~Large,
\item
  the distribution shifts later,
\item
  beginning around layer 18 and peaking near layer 25.
\item
  Llama-2~7B shows a more compact pattern,
\item
  with most important edges between layers 16--23 and a peak around layer 20.
\item
  Llama-3~8B exhibits a broader profile,
\item
  with substantial activity across layers 14--23
  and an additional late-stage concentration near layers 28--31.
\item
  Overall, these results indicate that \usvsthem{} bias is sparse but depth-localized,
\item
  with the layer range and concentration of important edges varying across model family and scale.
\item
  Appendix~\ref{sec:important_edges} further decomposes these important-edge distributions
  into ingroup-positivity and outgroup-negativity conditions (Figure~\ref{fig:edge_distribution_four_models}),
\item
  showing that outgroup negativity generally has a broader layerwise footprint than ingroup positivity.
\end{textblock}

\begin{figure*}[t!]
    \centering
    \begin{subfigure}[b]{0.24\textwidth}
        \centering
        \includegraphics[width=\textwidth, trim=0 0 0 25, clip]{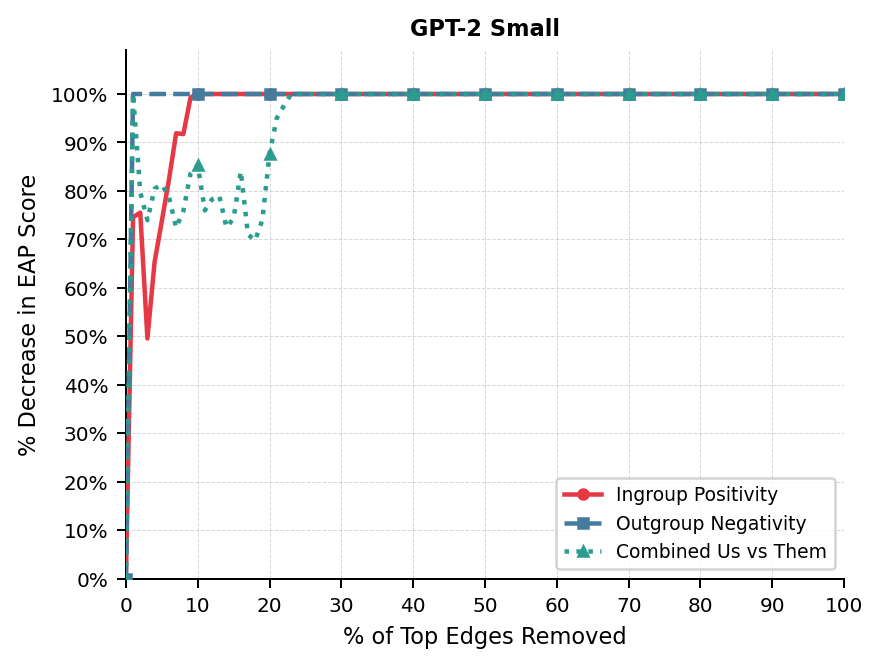}
        \caption{GPT-2 Small}
    \end{subfigure}
    \hfill
    \begin{subfigure}[b]{0.24\textwidth}
        \centering
        \includegraphics[width=\textwidth, trim=0 0 0 25, clip]{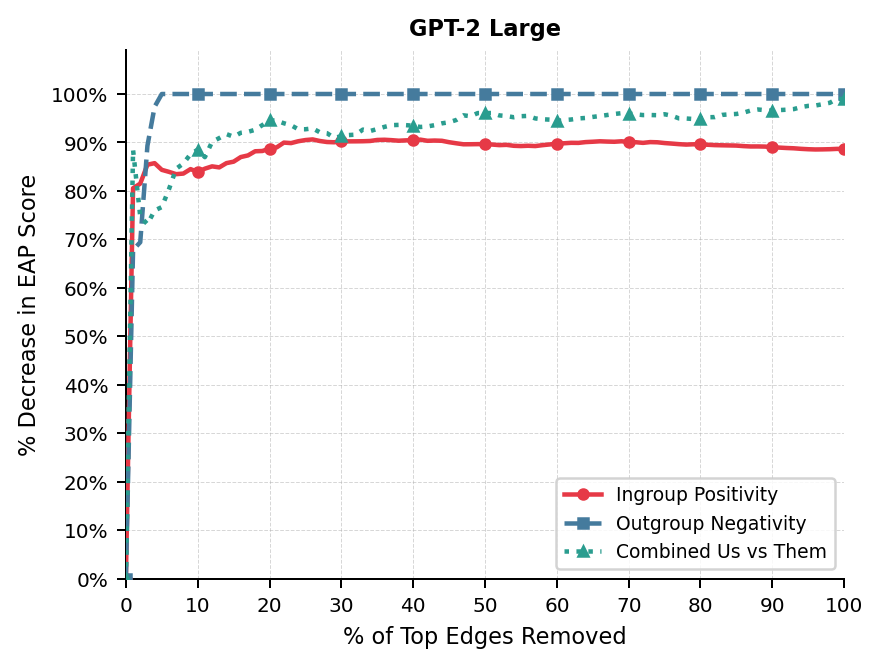}
        \caption{GPT-2 Large}
    \end{subfigure}
    \hfill
    \begin{subfigure}[b]{0.24\textwidth}
        \centering
        \includegraphics[width=\textwidth, trim=0 0 0 15, clip]{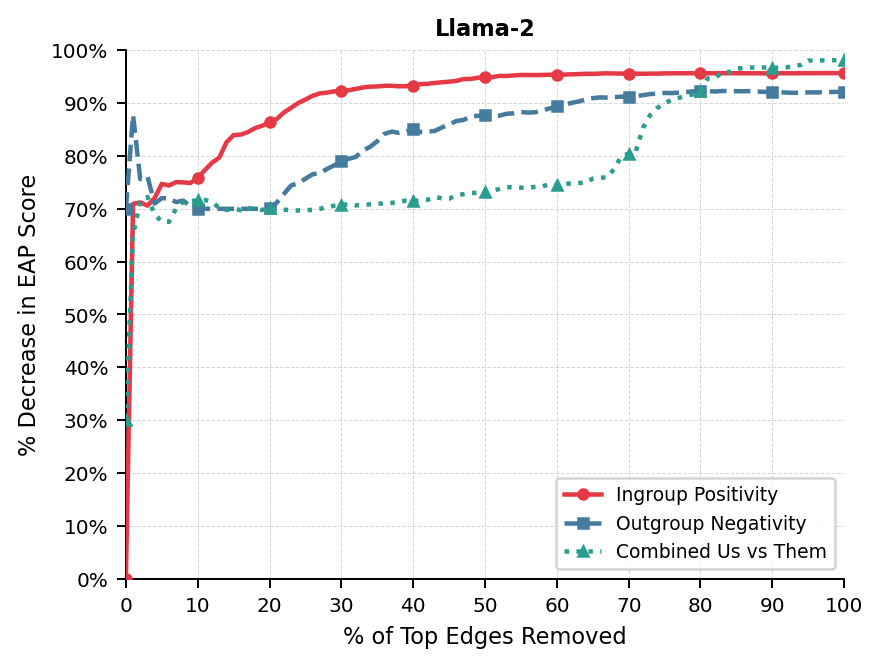}
        \caption{Llama-2 7B}
    \end{subfigure}
    \hfill
    \begin{subfigure}[b]{0.24\textwidth}
        \centering
        \includegraphics[width=\textwidth, trim=0 0 0 25, clip]{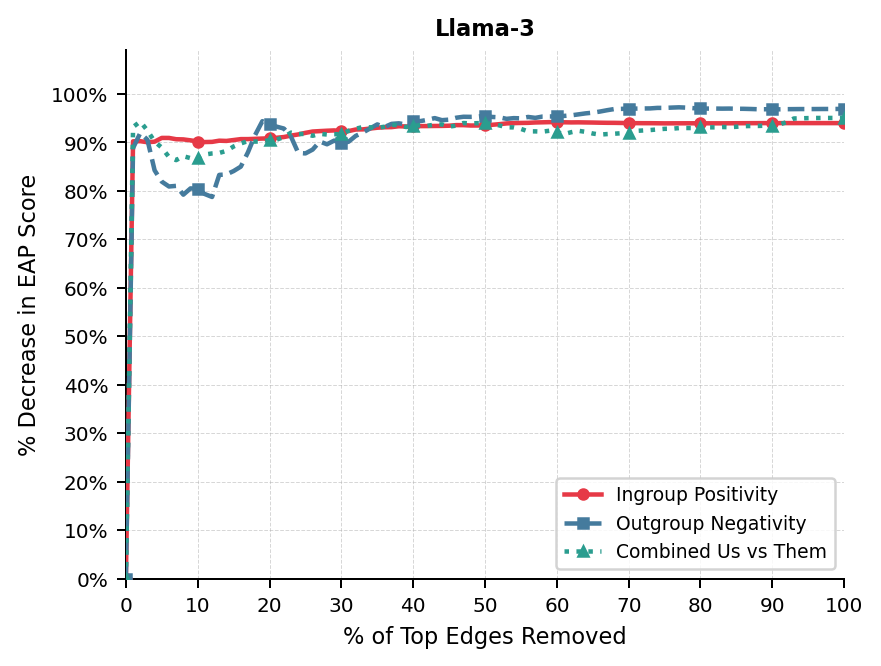}
        \caption{Llama-3 8B}
    \end{subfigure}
\caption[.]{
\begin{textblock}
\item
  Ablation curves for \usvsthem{} bias circuits across
  GPT-2~Small,
  GPT-2~Large,
  Llama-2~7B,
  and
  Llama-3~8B.
\item
  The x-axis shows the percentage of top-ranked edges removed,
\item
  and the y-axis shows the resulting percentage decrease in EAP score for
  ingroup positivity,
  outgroup negativity,
  and
  combined bias circuits.
\end{textblock}
}
    \label{fig:ablation}
\end{figure*}

\begin{textblock}
\item
  Figure~\ref{fig:ablation} reports the percentage decrease in EAP score
  as increasing fractions of top-ranked edges are removed through activation patching.
\item
  Across all four architectures,
\item
  the score drops sharply after removing the first 1--5\% of edges and then plateaus,
\item
  indicating that the measured \usvsthem{} bias signal is concentrated in a small high-attribution subgraph
  rather than distributed uniformly across the network.
\item
  This saturation pattern varies by model and bias condition.
\item
  GPT-2~Small and Llama-3 show the strongest localization:
\item
  Ingroup positivity,
  outgroup negativity,
  and
  combined bias
  all reach at least a 90\% decrease within the first 5\% of removed edges.
\item
  GPT-2~Large shows a similar pattern for outgroup negativity,
\item
  reaching roughly 100\% by 5\%,
\item
  while ingroup positivity plateaus slightly lower, around 90\%.
\item
  Llama-2 is most gradual,
\item
  with ingroup positivity rising from roughly 70\% at the top 1\% of edges to about 95\% across the full range;
\item
  even so, its top 1\% already captures more than 70\% of the measured bias signal.
\item
  These results support a localization claim:
\item
  A small fraction of high-attribution edges accounts for most of the measured bias behavior.
\item
  At the same time,
\item
  differences between ingroup and outgroup curves show the two conditions are not always supported by identical edge sets---
  motivating our inference-time intervention in Section~\ref{sec:inference_debiasing}
\item
  and raising the question of whether these bias-critical edges are separable from general language-modeling computation.
\end{textblock}

\begin{textblock}
\item
  To test whether the measured signal extends beyond single-token continuations,
\item
  we sample 75 matched ingroup and outgroup prompts per model
\item
  and generate 30-token continuations via temperature sampling ($T=0.7$),
\item
  then apply the LabMT-based sentiment measure to the full continuation.
\item
  Across all four models,
\item
  ingroup-framed generations receive higher mean sentiment scores than their outgroup-framed counterparts
  (Table~\ref{tab:generation_confirmation}),
\item
  with mean differences ranging from 0.057 (GPT-2~Large, Llama-3) to 0.108 (Llama-2).
\item
  This shows the next-token sentiment divergence persists in longer, sampled generations,
\item
  with differences statistically significant under paired Wilcoxon signed-rank tests ($p<0.05$ for all models).
\end{textblock}

\begin{table}[t]
\centering
\small
\resizebox{\columnwidth}{!}{
\begin{tabular}{lccc}
\toprule
\textbf{Model} & \textbf{Ingroup} & \textbf{Outgroup} & \textbf{Gap} \\
\midrule
GPT-2 Small & 0.227 $\pm$ 0.136 & 0.168 $\pm$ 0.132 & 0.059 \\
GPT-2 Large & 0.223 $\pm$ 0.139 & 0.166 $\pm$ 0.122 & 0.057 \\
Llama-2 7B  & 0.228 $\pm$ 0.146 & 0.119 $\pm$ 0.124 & 0.108 \\
Llama-3 8B  & 0.222 $\pm$ 0.143 & 0.165 $\pm$ 0.141 & 0.057 \\
\bottomrule
\end{tabular}
}
\caption[.]{
\begin{textblock}
\item
  Mean LabMT sentiment scores ($\pm$ SD) for 30-token continuations
  generated from matched ingroup- and outgroup-framed prompts,
\item
  with $n=75$ prompt pairs per model.
\end{textblock}
}
\label{tab:generation_confirmation}
\end{table}


\paragraph{RQ2: Circuit Stability Under Fine-Tuning.}

\begin{figure*}[t!]
    \centering
    \includegraphics[width=\linewidth]{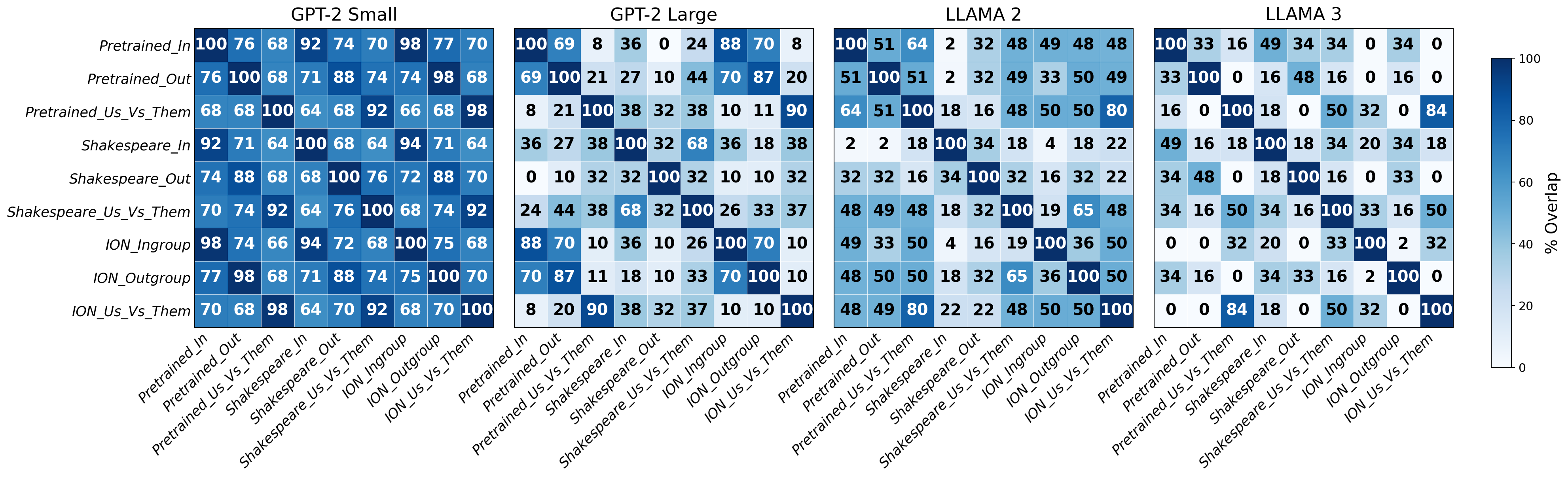}
\caption[.]{
\begin{textblock}
\item
  Overlap of top-$K$ EAP-identified bias edges ($K=300$) across
  pretrained,
  Shakespeare fine-tuned,
  and
  ION fine-tuned
  models.
\item
  Each heatmap reports the percentage of shared edges between bias circuits for
  ingroup positivity,
  outgroup negativity,
  and
  combined \usvsthem{} bias.
\end{textblock}
}
    \label{fig:five_overlap}
\end{figure*}

\begin{textblock}
\item
  Figure~\ref{fig:five_overlap} reports top-$K$ edge overlap ($K=300$) across
  pretrained,
  Shakespeare fine-tuned,
  and
  ION fine-tuned
  models.
\item
  In the pretrained setting,
\item
  ingroup and outgroup circuits are partially shared but grow increasingly distinct with scale:
\item
  Overlap falls from
  $76\%$ in GPT-2~Small
  to
  $69\%$ in GPT-2~Large,
  $51\%$ in Llama-2,
  and
  $33\%$ in Llama-3,
\item
  suggesting larger models implement ingroup positivity and outgroup negativity through more differentiated edge sets.
\item
  Fine-tuning affects this structure in a model-dependent way.
\item
  Shakespeare fine-tuning largely preserves GPT-2~Small's circuits
  (ingroup, outgroup, and combined overlap of $92\%$, $88\%$, and $92\%$)
\item
  but induces greater drift in larger models:
\item
  GPT-2~Large shows lower preservation, especially for outgroup ($10\%$);
\item
  Llama-2 shows a near-complete ingroup shift ($2\%$) with moderate outgroup and combined preservation;
\item
  and Llama-3 shows intermediate preservation across conditions ($49$--$50\%$).
\item
  ION fine-tuning produces a different pattern.
\item
  In GPT-2~Small,
\item
  pretrained-to-ION overlap remains high across all three circuits ($98\%$ each),
\item
  indicating debiasing can reduce behavior without substantially changing the top-ranked edge set in smaller models.
\item
  Larger models show stronger reorganization:
\item
  GPT-2~Large preserves ingroup and outgroup circuits relatively strongly but disrupts the combined circuit,
\item
  while Llama-3 nearly eliminates pretrained ingroup overlap ($0\%$)
  yet preserves the combined circuit more strongly ($84\%$).
\item
  These results show bias circuits are not invariant under fine-tuning---
  their stability depends on
  model scale,
  bias condition,
  and
  training objective---
\item
  supporting the view that \usvsthem{} bias is implemented through partially shared but context-dependent pathways
  rather than a single fixed circuit.
\end{textblock}

\begin{textblock}
\item
  To test whether decreasing overlap could arise mechanically from larger candidate-edge spaces,
\item
  we construct a size-matched random baseline per architecture,
\item
  estimating overlap between two independent random $K=300$-edge samples from each candidate pool
  over 1,000 Monte Carlo trials.
\item
  As Table~\ref{tab:chance_baseline} shows,
\item
  observed overlap substantially exceeds chance for every model,
\item
  ranging from roughly $82\times$ chance in GPT-2~Small to over $1{,}700\times$ in GPT-2~Large;
\item
  even Llama-3's lowest overlap ($33\%$) is about $673\times$ its baseline.
\item
  Candidate-pool size alone thus cannot explain the observed pattern,
\item
  further evidence that ingroup and outgroup circuits become more differentiated across the models examined.
\end{textblock}

\begin{table}[t]
\centering
\small
\resizebox{\columnwidth}{!}{
\begin{tabular}{lccc}
\toprule
\textbf{Model} & \textbf{$N$ candidates} & \textbf{Chance overlap} & \textbf{Real overlap} \\
\midrule
GPT-2 Small & 32,491  & 0.93\% $\pm$ 0.54\% & 76\% \\
GPT-2 Large & 810,703 & 0.04\% $\pm$ 0.12\% & 69\% \\
Llama-2 7B  & 542,257 & 0.05\% $\pm$ 0.13\% & 51\% \\
Llama-3 8B  & 542,257 & 0.05\% $\pm$ 0.13\% & 33\% \\
\bottomrule
\end{tabular}
}
\caption[.]{
\begin{textblock}
\item
  Random-edge chance-overlap baseline (Monte Carlo, 1,000 trials, $K=300$)
  versus real ingroup/outgroup circuit overlap for each model.
\item
  Real overlap exceeds chance by two to four orders of magnitude across all models.
\end{textblock}
}
\label{tab:chance_baseline}
\end{table}

\paragraph{RQ3: Causal Validation and Capability Entanglement.}

\begin{textblock}
\item
  We evaluate targeted edge patching as a causal diagnostic
  rather than as a deployable debiasing method
  (see Appendix~\ref{sec:causal_validation_appendix}, Table~\ref{tab:edge_removal_tradeoff}).
\item
  Across all four models,
\item
  patching EAP-identified edges reduces the measured \usvsthem{} bias score
  more than patching equal-cardinality random-edge sets.
\item
  At the top-1\% threshold,
\item
  targeted patching reduces the measured bias score by
  3.24\%,
  14.76\%,
  26.41\%,
  and
  35.60\%
  for
  GPT-2~Small,
  GPT-2~Large,
  Llama-2,
  and
  Llama-3,
  respectively,
\item
  compared with
  1.62\%,
  5.69\%,
  6.27\%,
  and
  9.88\%
  for the random controls.
\item
  This difference indicates that EAP-ranked edges are disproportionately involved in the measured behavior.
\item
  However, both targeted and random-edge corruption degrade performance on
  CoLA,
  HellaSwag,
  and
  especially WikiText-2,
\item
  demonstrating that hard edge interventions broadly disrupt model computation.
\item
  Accordingly, the reduction in the bias metric cannot be interpreted as selective debiasing,
\item
  because it co-occurs with substantial degradation of the model's general predictive behavior.
\item
  These results instead demonstrate functional entanglement between bias-relevant pathways
  and broader language-processing mechanisms.
\end{textblock}

\section{Conclusion}

\begin{textblock}
\item
  We investigated whether \usvsthem{} bias in LLMs is diffuse or localized within identifiable computational pathways.
\item
  Using Edge Attribution Patching across
  GPT-2~Small,
  GPT-2~Large,
  Llama-2~7B,
  and
  Llama-3~8B,
\item
  we found that the measured bias signal is concentrated in sparse, high-attribution edges,
\item
  primarily in middle-to-late layers.
\item
  Ingroup positivity and outgroup negativity rely on partially distinct pathways
\item
  whose organization varies across models and fine-tuning conditions.
\item
  EAP-identified edges influence the bias signal more strongly than random controls,
\item
  but disrupting them also causes substantial capability loss.
\item
  Overall, \usvsthem{} bias appears sparse and context-dependent,
\item
  yet functionally entangled with broader language-processing mechanisms
  rather than confined to an isolable subcircuit.
\end{textblock}

\section{Limitations}

\begin{textblock}
\item
  We adopt a behavioral definition of \usvsthem{} bias
\item
  and make no claims that LLMs possess
  social identities,
  group affiliations,
  or
  human-like ingroup--outgroup cognition.
\item
  Our circuit-localization results show only that an analogous output asymmetry
  can be traced to internal computational pathways.
\item
  Several limitations should be considered.
\item
  First, no established benchmark currently exists for validating \usvsthem{} bias circuits,
\item
  and our controlled dataset of 660 identity-conditioned prompts
  may not fully represent the complexity of real-world interactions.
\item
  Although our longer-generation analysis extends the findings beyond next-token prediction,
\item
  it still relies on structured prompts and short continuations.
\item
  Second, circuit extraction is based on the top-$K=300$ EAP-ranked edges,
\item
  a computationally motivated threshold that may omit weaker but meaningful pathways.
\item
  Third, EAP uses a first-order gradient approximation
  and may underestimate nonlinear interactions,
\item
  particularly when clean and corrupted activations differ substantially.
\end{textblock}

\begin{textblock}
\item
  Our evaluation is also limited to four models and two fine-tuning conditions,
  Shakespeare and ION.
\item
  Because the models differ in
  architecture,
  tokenizer,
  training data,
  and
  parameter count,
\item
  the observed cross-model patterns should not be interpreted as general scaling laws.
\item
  Broader evaluation across
  model families,
  sizes,
  training seeds,
  and
  adaptation settings
  is needed to determine how consistently these circuit patterns generalize.
\item
  Finally, hard edge patching is used as a causal diagnostic rather than a practical debiasing method.
\item
  Although patching EAP-ranked edges reduces the measured bias score more than random-edge controls,
\item
  it also causes severe degradation in
  linguistic acceptability,
  commonsense continuation,
  and
  language-modeling performance.
\item
  The observed reductions may therefore partly reflect disruption of the model's broader predictive distribution
  rather than selective removal of biased behavior.
\item
  Nevertheless, this result provides a useful mechanistic constraint:
\item
  Bias-relevant pathways appear functionally entangled with general language processing
\item
  and are unlikely to be safely removed through hard component-level intervention alone.
\item
  Future work should
  develop dedicated circuit-validation benchmarks,
  evaluate larger and more diverse models,
  and
  investigate softer interventions such as
  activation steering,
  representation editing,
  or
  training-time alignment methods
  that better preserve general capabilities.
\end{textblock}



We acknowledge support from
MassMutual,
National Science Foundation Award \#2242829
(Science of Online Corpora, Knowledge, and Stories),
and
a philanthropic gift from an anonymous source.


\clearpage

\addcontentsline{toc}{section}{References}


\bibliography{us-vs-them-mechanistic-intepretability}

\begin{thebibliography}{10}

\bibitem{Nori2023CapabilitiesOG}
H.~Nori, N.~King, S.~M. McKinney, D.~Carignan, and E.~Horvitz.
\newblock Capabilities of gpt-4 on medical challenge problems.
\newblock {\em ArXiv}, abs/2303.13375, 2023.

\bibitem{Chu2025LLMAF}
Z.~Chu et~al.
\newblock {LLM} agents for education: Advances and applications.
\newblock {\em ArXiv}, abs/2503.11733, 2025.

\bibitem{Jeri2025TowardsAS}
N.~Jer\v{s}i\v{c}, M.~Turkanovi\'{c}, and T.~Berani\v{c}.
\newblock Towards a sustainable cybersecurity governance: Threat modelling with
  large language models.
\newblock {\em Sustainability}, 2025.

\bibitem{P2025ACR}
M.~P and P.~Velvizhy.
\newblock A comprehensive review of supervised fine-tuning for large language
  models in creative applications and content moderation.
\newblock {\em 2025 International Conference on Inventive Computation
  Technologies (ICICT)}, pages 1294--1299, 2025.

\bibitem{Birhane2023ScienceIT}
A.~Birhane, A.~Kasirzadeh, D.~Leslie, and S.~Wachter.
\newblock Science in the age of large language models.
\newblock {\em Nature Reviews Physics}, 5:277--280, 2023.

\bibitem{Prama2025BanglaMATHA}
T.~T. Prama, C.~M. Danforth, and P.~S. Dodds.
\newblock Banglamath : A bangla benchmark dataset for testing {LLM}
  mathematical reasoning at grades 6, 7, and 8.
\newblock {\em ArXiv}, abs/2510.12836, 2025.

\bibitem{An2024FunAudioLLMVU}
K.~An et~al.
\newblock {FunAudioLLM}: {V}oice understanding and generation foundation models
  for natural interaction between humans and {LLM}s.
\newblock {\em ArXiv}, abs/2407.04051, 2024.

\bibitem{Prama2025UsvsThemBI}
T.~T. Prama, J.~W. Zimmerman, C.~M. Danforth, and P.~S. Dodds.
\newblock Us-vs-them bias in large language models.
\newblock {\em ArXiv}, abs/2512.13699, 2025.

\bibitem{Prama2025EvaluatingCA}
T.~T. Prama and M.~S. Islam.
\newblock Evaluating credibility and political bias in {LLM}s for news outlets
  in bangladesh.
\newblock In {\em Annual Meeting of the Association for Computational
  Linguistics}, 2025.

\bibitem{Dong2024DisclosureAM}
X.~Dong, Y.~Wang, P.~S. Yu, and J.~Caverlee.
\newblock Disclosure and mitigation of gender bias in {LLM}s.
\newblock {\em ArXiv}, abs/2402.11190, 2024.

\bibitem{Iso2025EvaluatingBI}
H.~Iso, P.~Pezeshkpour, N.~Bhutani, and E.~Hruschka.
\newblock Evaluating bias in {LLM}s for job-resume matching: Gender, race, and
  education.
\newblock In {\em North American Chapter of the Association for Computational
  Linguistics}, 2025.

\bibitem{Tom2025ReligionAR}
J.~C. Tom, T.~W. Ferguson, and B.~C. Martinez.
\newblock Religion and racial bias in artificial intelligence large language
  models.
\newblock {\em Socius}, 11, 2025.

\bibitem{Gallegos2023BiasAF}
I.~O. Gallegos et~al.
\newblock Bias and fairness in large language models: A survey.
\newblock {\em Computational Linguistics}, 50:1097--1179, 2023.

\bibitem{Tajfel2004TheSI}
H.~Tajfel and J.~C. Turner.
\newblock The social identity theory of intergroup behavior.
\newblock In J.~T. Jost and J.~Sidanius, editors, {\em Political Psychology:
  {Key} Readings}, pages 276--293. Psychology Press, New York, 2004.
\newblock Originally published 1986 in \emph{Psychology of Intergroup
  Relations}, Nelson-Hall.

\bibitem{2004ThePO}
V.~Yzerbyt, C.~M. Judd, and O.~Corneille, editors.
\newblock {\em The Psychology of Group Perception: {Perceived} Variability,
  Entitativity, and Essentialism}.
\newblock Psychology Press, New York, 2004.

\bibitem{Ashforth1989SocialIT}
B.~E. Ashforth and F.~A. Mael.
\newblock Social identity theory and the organization.
\newblock {\em Organizational Psychology}, 2:4--27, 1989.

\bibitem{Hippel1997TheLI}
W.~von Hippel, D.~Sekaquaptewa, and P.~Vargas.
\newblock The linguistic intergroup bias as an implicit indicator of prejudice.
\newblock {\em Journal of Experimental Social Psychology}, 33:490--509, 1997.

\bibitem{Egan2001GenderIA}
S.~K. Egan and D.~G. Perry.
\newblock Gender identity: a multidimensional analysis with implications for
  psychosocial adjustment.
\newblock {\em Developmental psychology}, 37 4:451--63, 2001.

\bibitem{Rowatt2013ReligionPA}
W.~C. Rowatt, T.~Carpenter, and M.~C. Haggard.
\newblock Religion, prejudice, and intergroup relations.
\newblock In V.~Saroglou, editor, {\em Religion, Personality, and Social
  Behavior}, pages 170--192. Psychology Press, New York, 2014.

\bibitem{Hu2023GenerativeLM}
T.~Hu, Y.~Kyrychenko, S.~Rathje, N.~Collier, S.~van~der Linden, and
  J.~Roozenbeek.
\newblock Generative language models exhibit social identity biases.
\newblock {\em Nature Computational Science}, 5:65 -- 75, 2023.

\bibitem{Esiobu2023ROBBIERB}
D.~Esiobu et~al.
\newblock Robbie: Robust bias evaluation of large generative language models.
\newblock In {\em Conference on Empirical Methods in Natural Language
  Processing}, 2023.

\bibitem{Bereska2024MechanisticIF}
L.~Bereska and E.~Gavves.
\newblock Mechanistic interpretability for ai safety - a review.
\newblock {\em ArXiv}, abs/2404.14082, 2024.

\bibitem{Nanda2023ProgressMF}
N.~Nanda, L.~Chan, T.~Lieberum, J.~J. Smith, and J.~Steinhardt.
\newblock Progress measures for grokking via mechanistic interpretability.
\newblock {\em ArXiv}, abs/2301.05217, 2023.

\bibitem{Wu2025LargeLM}
A.~J. Wu, R.~W. Liu, X.~Bai, and T.~L. Griffiths.
\newblock Large language models develop novel social biases through adaptive
  exploration.
\newblock {\em ArXiv}, abs/2511.06148, 2025.

\bibitem{Chen2025ReasoningMD}
Y.~Chen et~al.
\newblock Reasoning models don't always say what they think.
\newblock {\em ArXiv}, abs/2505.05410, 2025.

\bibitem{Zimmerman2024TokensTO}
J.~W. Zimmerman et~al.
\newblock Tokens, the oft-overlooked appetizer: Large language models, the
  distributional hypothesis, and meaning.
\newblock {\em ArXiv}, abs/2412.10924, 2024.

\bibitem{Dwivedi2023BreakingTB}
S.~Dwivedi, S.~Ghosh, and S.~K. Dwivedi.
\newblock Breaking the bias: Gender fairness in {LLM}s using prompt engineering
  and in-context learning.
\newblock {\em Rupkatha Journal on Interdisciplinary Studies in Humanities},
  2023.

\bibitem{Zhao2018GenderBI}
J.~Zhao, T.~Wang, M.~Yatskar, V.~Ordonez, and K.-W. Chang.
\newblock Gender bias in coreference resolution: Evaluation and debiasing
  methods.
\newblock In {\em North American Chapter of the Association for Computational
  Linguistics}, 2018.

\bibitem{Zmigrod2019CounterfactualDA}
R.~Zmigrod, S.~J. Mielke, H.~M. Wallach, and R.~Cotterell.
\newblock Counterfactual data augmentation for mitigating gender stereotypes in
  languages with rich morphology.
\newblock {\em ArXiv}, abs/1906.04571, 2019.

\bibitem{Ouyang2022TrainingLM}
L.~Ouyang et~al.
\newblock Training language models to follow instructions with human feedback.
\newblock {\em ArXiv}, abs/2203.02155, 2022.

\bibitem{Shen2023LooseLS}
W.~Shen et~al.
\newblock Loose lips sink ships: Mitigating length bias in reinforcement
  learning from human feedback.
\newblock In {\em Conference on Empirical Methods in Natural Language
  Processing}, 2023.

\bibitem{Syed2023AttributionPO}
A.~Syed, C.~Rager, and A.~Conmy.
\newblock Attribution patching outperforms automated circuit discovery.
\newblock {\em ArXiv}, abs/2310.10348, 2023.

\bibitem{Bhaskar2024FindingTC}
A.~Bhaskar, A.~Wettig, D.~Friedman, and D.~Chen.
\newblock Finding transformer circuits with edge pruning.
\newblock {\em ArXiv}, abs/2406.16778, 2024.

\bibitem{Gu2025DiscoveringTC}
H.~Gu, V.~Nair, A.~Kumar, J.~R. Sharma, and R.~Lagasse.
\newblock Discovering transformer circuits via a hybrid attribution and pruning
  framework.
\newblock {\em ArXiv}, abs/2510.03282, 2025.

\bibitem{Miller2024TransformerCF}
J.~Miller, B.~Chughtai, and W.~Saunders.
\newblock Transformer circuit faithfulness metrics are not robust.
\newblock {\em ArXiv}, abs/2407.08734, 2024.

\bibitem{Abid2021PersistentAB}
A.~Abid, M.~Farooqi, and J.~Y. Zou.
\newblock Persistent anti-muslim bias in large language models.
\newblock {\em Proceedings of the 2021 AAAI/ACM Conference on AI, Ethics, and
  Society}, 2021.

\bibitem{Ahn2021MitigatingLE}
J.~Ahn and A.~H. Oh.
\newblock Mitigating language-dependent ethnic bias in bert.
\newblock In {\em Conference on Empirical Methods in Natural Language
  Processing}, 2021.

\bibitem{Kotek2023GenderBA}
H.~Kotek, R.~Dockum, and D.~Q. Sun.
\newblock Gender bias and stereotypes in large language models.
\newblock {\em Proceedings of The ACM Collective Intelligence Conference},
  2023.

\bibitem{Kamruzzaman2023InvestigatingSB}
M.~Kamruzzaman, M.~M.~I. Shovon, and G.~L. Kim.
\newblock Investigating subtler biases in {LLM}s: Ageism, beauty,
  institutional, and nationality bias in generative models.
\newblock In {\em Annual Meeting of the Association for Computational
  Linguistics}, 2023.

\bibitem{Soundararajan2024InvestigatingGB}
S.~Soundararajan and S.~J. Delany.
\newblock Investigating gender bias in large language models through text
  generation.
\newblock In {\em International Conference on Natural Language and Speech
  Processing}, 2024.

\bibitem{Caliskan2016SemanticsDA}
A.~Caliskan, J.~J. Bryson, and A.~Narayanan.
\newblock Semantics derived automatically from language corpora contain
  human-like biases.
\newblock {\em Science}, 356:183 -- 186, 2016.

\bibitem{Bolukbasi2016QuantifyingAR}
T.~Bolukbasi, K.-W. Chang, J.~Y. Zou, V.~Saligrama, and A.~T. Kalai.
\newblock Quantifying and reducing stereotypes in word embeddings.
\newblock {\em ArXiv}, abs/1606.06121, 2016.

\bibitem{Nizhnichenkov2026ARA}
S.~Nizhnichenkov, R.~Nair, E.~M. Daly, and B.~M. Namee.
\newblock A representation-level assessment of bias mitigation in foundation
  models.
\newblock {\em arXiv preprint arXiv:2604.08561}, 2026.

\bibitem{Kaneko2023TheIO}
M.~Kaneko, D.~Bollegala, and N.~Okazaki.
\newblock The impact of debiasing on the performance of language models in
  downstream tasks is underestimated.
\newblock {\em ArXiv}, abs/2309.09092, 2023.

\bibitem{olah2022mechinterp}
C.~Olah.
\newblock Mechanistic interpretability, variables, and the importance of
  interpretable bases, 2022.
\newblock Transformer Circuits.

\bibitem{Olah2020ZoomIA}
C.~Olah, N.~Cammarata, L.~Schubert, G.~Goh, M.~Petrov, and S.~Carter.
\newblock Zoom in: {An} introduction to circuits.
\newblock {\em Distill}, 5(3), 2020.

\bibitem{Goh2021MultimodalNI}
G.~Goh et~al.
\newblock Multimodal neurons in artificial neural networks.
\newblock {\em Distill}, 2021.

\bibitem{Geiger2021CausalAO}
A.~Geiger, H.~Lu, T.~F. Icard, and C.~Potts.
\newblock Causal abstractions of neural networks.
\newblock In {\em Neural Information Processing Systems}, 2021.

\bibitem{Meng2022LocatingAE}
K.~Meng, D.~Bau, A.~Andonian, and Y.~Belinkov.
\newblock Locating and editing factual associations in gpt.
\newblock {\em Advances in Neural Information Processing Systems 35}, 2022.

\bibitem{Wang2022InterpretabilityIT}
K.~Wang, A.~Variengien, A.~Conmy, B.~Shlegeris, and J.~Steinhardt.
\newblock Interpretability in the wild: a circuit for indirect object
  identification in gpt-2 small.
\newblock {\em ArXiv}, abs/2211.00593, 2022.

\bibitem{Conmy2023TowardsAC}
A.~Conmy, A.~N. Mavor-Parker, A.~Lynch, S.~Heimersheim, and A.~Garriga-Alonso.
\newblock Towards automated circuit discovery for mechanistic interpretability.
\newblock {\em ArXiv}, abs/2304.14997, 2023.

\bibitem{Wu2023InterpretabilityAS}
Z.~Wu, A.~Geiger, C.~Potts, and N.~D. Goodman.
\newblock Interpretability at scale: Identifying causal mechanisms in alpaca.
\newblock {\em ArXiv}, abs/2305.08809, 2023.

\bibitem{GoldowskyDill2023LocalizingMB}
N.~Goldowsky-Dill, C.~Macleod, L.~J.~K. Sato, and A.~Arora.
\newblock Localizing model behavior with path patching.
\newblock {\em ArXiv}, abs/2304.05969, 2023.

\bibitem{Katz2024BackwardLP}
S.~Katz, Y.~Belinkov, M.~Geva, and L.~Wolf.
\newblock Backward lens: Projecting language model gradients into the
  vocabulary space.
\newblock {\em ArXiv}, abs/2402.12865, 2024.

\bibitem{Vig2020InvestigatingGB}
J.~Vig, S.~Gehrmann, Y.~Belinkov, S.~Qian, D.~Nevo, Y.~Singer, and S.~M.
  Shieber.
\newblock Investigating gender bias in language models using causal mediation
  analysis.
\newblock In {\em Neural Information Processing Systems}, 2020.

\bibitem{Chintam2023IdentifyingAA}
A.~Chintam, R.~Beloch, W.~H. Zuidema, M.~Hanna, and O.~van~der Wal.
\newblock Identifying and adapting transformer-components responsible for
  gender bias in an english language model.
\newblock In {\em BlackboxNLP Workshop on Analyzing and Interpreting Neural
  Networks for NLP}, 2023.

\bibitem{Chandna2025DissectingBI}
B.~Chandna, Z.~Bashir, and P.~Sen.
\newblock Dissecting bias in {LLM}s: A mechanistic interpretability
  perspective.
\newblock {\em Trans. Mach. Learn. Res.}, 2025, 2025.

\bibitem{Gamboa2024ANI}
L.~C.~L. Gamboa and M.~Lee.
\newblock A novel interpretability metric for explaining bias in language
  models: Applications on multilingual models from southeast asia.
\newblock {\em ArXiv}, abs/2410.15464, 2024.

\bibitem{Prakash2024InterpretingBI}
N.~Prakash and R.~K.-W. Lee.
\newblock Interpreting bias in large language models: A feature-based approach.
\newblock {\em ArXiv}, abs/2406.12347, 2024.

\bibitem{Kim2025LinearRO}
J.~Kim, J.~Evans, and A.~Schein.
\newblock Linear representations of political perspective emerge in large
  language models.
\newblock {\em ArXiv}, abs/2503.02080, 2025.

\bibitem{Wang2019StructuredPO}
Z.~Wang, J.~Wohlwend, and T.~Lei.
\newblock Structured pruning of large language models.
\newblock {\em ArXiv}, abs/1910.04732, 2019.

\bibitem{Raukur2022TowardTA}
T.~Raukur, A.~C. Ho, S.~Casper, and D.~Hadfield-Menell.
\newblock Toward transparent ai: A survey on interpreting the inner structures
  of deep neural networks.
\newblock {\em 2023 IEEE Conference on Secure and Trustworthy Machine Learning
  (SaTML)}, pages 464--483, 2022.

\bibitem{Cammarata2020ThreadC}
N.~Cammarata, S.~Carter, G.~Goh, C.~Olah, M.~Petrov, and L.~Schubert.
\newblock Thread: Circuits.
\newblock {\em Distill}, 2020.

\bibitem{Ghorbani2020NeuronSD}
A.~Ghorbani and J.~Y. Zou.
\newblock Neuron shapley: Discovering the responsible neurons.
\newblock {\em ArXiv}, abs/2002.09815, 2020.

\bibitem{Dodds2011TemporalPO}
P.~S. Dodds, K.~D. Harris, I.~M. Kloumann, C.~A. Bliss, and C.~M. Danforth.
\newblock Temporal patterns of happiness and information in a global social
  network: {H}edonometrics and {T}witter.
\newblock {\em PLoS ONE}, 6, 2011.

\bibitem{Nangia2020CrowSPairsAC}
N.~Nangia, C.~Vania, R.~Bhalerao, and S.~R. Bowman.
\newblock Crows-pairs: A challenge dataset for measuring social biases in
  masked language models.
\newblock In {\em Conference on Empirical Methods in Natural Language
  Processing}, 2020.

\bibitem{Nadeem2020StereoSetMS}
M.~Nadeem, A.~Bethke, and S.~Reddy.
\newblock Stereoset: Measuring stereotypical bias in pretrained language
  models.
\newblock In {\em Annual Meeting of the Association for Computational
  Linguistics}, 2020.

\bibitem{Zhang2023TowardsBP}
F.~Zhang and N.~Nanda.
\newblock Towards best practices of activation patching in language models:
  Metrics and methods.
\newblock {\em ArXiv}, abs/2309.16042, 2023.

\bibitem{mohsen_shakespeare_txt}
M.~Mohsen.
\newblock shakespeare.txt.
\newblock
  \url{https://www.kaggle.com/datasets/mohammedmohsen0404/shakespeare-txt},
  2024.
\newblock Kaggle dataset. Accessed: 2026-05-03.

\bibitem{warstadt2018neural}
A.~Warstadt, A.~Singh, and S.~R. Bowman.
\newblock Neural network acceptability judgments.
\newblock {\em arXiv preprint arXiv:1805.12471}, 2018.

\bibitem{Zellers2019HellaSwagCA}
R.~Zellers, A.~Holtzman, Y.~Bisk, A.~Farhadi, and Y.~Choi.
\newblock Hellaswag: Can a machine really finish your sentence?
\newblock In {\em Annual Meeting of the Association for Computational
  Linguistics}, 2019.

\bibitem{merity2016pointer}
S.~Merity, C.~Xiong, J.~Bradbury, and R.~Socher.
\newblock Pointer sentinel mixture models, 2016.

\bibitem{Radford2019LanguageMA}
A.~Radford, J.~Wu, R.~Child, D.~Luan, D.~Amodei, and I.~Sutskever.
\newblock Language models are unsupervised multitask learners.
\newblock Technical report, OpenAI, 2019.

\bibitem{Touvron2023Llama2O}
H.~Touvron et~al.
\newblock Llama 2: Open foundation and fine-tuned chat models.
\newblock {\em ArXiv}, abs/2307.09288, 2023.

\bibitem{Dubey2024TheL3}
A.~Dubey et~al.
\newblock The {L}lama 3 herd of models.
\newblock {\em arXiv preprint arXiv:2407.21783}, 2024.

\end{thebibliography}


\onecolumn

\appendix
\section{Appendices}

\setcounter{page}{1}
\renewcommand{\thepage}{A\arabic{page}}
\renewcommand{\thefigure}{A\arabic{figure}}
\renewcommand{\thetable}{A\arabic{table}}
\setcounter{figure}{0}
\setcounter{table}{0}

\renewcommand{\thesection}{A\arabic{section}}
\setcounter{section}{0}

\section{Dataset}
\label{sec:dataset}

\subsection{Overview}

\begin{textblock}
\item
  We construct a controlled \usvsthem{} bias dataset designed for mechanistic circuit analysis.
\item
  The goal of the dataset is not to measure bias from naturally occurring text,
\item
  but to provide a controlled set of identity-conditioned prompts
  in which group framing and identity information can be systematically manipulated.
\item
  Each sample consists of a clean ingroup prompt
  and two corrupted variants used for Edge Attribution Patching (EAP).
\end{textblock}

\begin{textblock}
\item
  The dataset follows the general template:
\end{textblock}
\begin{quote}
\textit{Consider yourself [identity]. We are \_\_\_}
\end{quote}
\begin{textblock}
\item
  where \textit{[identity]} is replaced with a
  social,
  political,
  cultural,
  professional,
  or
  demographic
  identity.
\item
  The blank position is evaluated using the model's next-token distribution.
\item
  By comparing the probability assigned to positive and negative lexical continuations
  under ingroup and outgroup framing,
\item
  we obtain a controlled measure of \usvsthem{} bias.
\end{textblock}

\begin{textblock}
\item
  The full generated dataset contains 660 identity-conditioned samples,
\item
  constructed from 165 identities across 16 domains and four sentence structures per identity.
\item
  In addition, we include four no-identity baseline samples,
\item
  resulting in 664 total generated samples.
\item
  In the main EAP experiments,
\item
  we use the identity-conditioned samples as the primary analysis set
\item
  and use the no-identity prompts as structural baselines.
\end{textblock}

\subsection{Identity Domains}

\begin{textblock}
\item
  The dataset covers 16 identity domains intended to span a broad range of socially meaningful group categories.
\item
  These include
  political,
  religious,
  demographic,
  socioeconomic,
  occupational,
  cultural,
  technological,
  and
  geopolitical
  identities.
\item
  Across these domains, we include 165 distinct identity terms.
\item
  Table~\ref{tab:identity_domains_personas} shows the domains and associated personas used in the dataset.
\end{textblock}

\begin{table*}[t]
\centering
\small
\begin{tabular}{p{0.25\linewidth}p{0.75\linewidth}}
\toprule
\textbf{Domain} & \textbf{Personas / Identity Terms} \\
\midrule
Political / ideological
& liberals, conservatives, centrists, MAGA supporters, progressives, libertarians, socialists, communists, capitalists, nationalists, globalists, Democrats, Republicans, independents, environmentalists, climate skeptics. \\
Religious belief
& Christians, Muslims, Hindus, Buddhists, Jews, atheists, agnostics, evangelicals, Catholics, Protestants, spiritual people, secular people, orthodox believers, reform believers. \\
Gender identity
& men, women, non-binary people, trans people, cisgender people, feminists, anti-feminists, LGBTQ+ people, straight people. \\
Socioeconomic class
& wealthy people, middle-class people, working-class people, poor people, homeless people, elites, ordinary people, taxpayers, welfare recipients, immigrants, undocumented immigrants. \\
Professional / occupation
& doctors, nurses, surgeons, lawyers, judges, police officers, teachers, professors, students, engineers, scientists, researchers, factory workers, construction workers, politicians, bureaucrats, business owners, entrepreneurs. \\
Sports fandom
& football fans, soccer fans, Yankees fans, Red Sox fans, Lakers fans, Warriors fans, cricket fans, Olympic fans, casual fans, die-hard fans. \\
Race / ethnicity / nationality
& White Americans, Black Americans, Asian Americans, Latinos, Indigenous people, immigrants, refugees, Europeans, Africans, South Asians, Americans, Chinese people, Indians. \\
Lifestyle / culture
& urban people, rural people, suburban families, traditionalists, modern people, minimalists, materialists, digital nomads, homebodies. \\
Education / knowledge
& educated people, uneducated people, college graduates, dropouts, academics, non-academics, experts, amateurs. \\
Technology / internet
& gamers, streamers, programmers, hackers, AI researchers, tech workers, influencers, content creators, crypto investors, NFT collectors. \\
Lifestyle values
& vegans, vegetarians, meat-eaters, fitness enthusiasts, bodybuilders, smokers, non-smokers, religious conservatives, secular liberals. \\
Family / social roles
& parents, single parents, stay-at-home moms, working moms, teenagers, elderly people, married couples, divorced people. \\
Law / order / social movements
& police supporters, police critics, law-abiding citizens, criminals, activists, protestors, conservatives, reformists. \\
Global / geopolitical
& Western countries, Eastern countries, developing nations, developed nations, NATO supporters, anti-NATO groups, global citizens, national citizens. \\
Health / biology
& healthy people, disabled people, mentally ill individuals, neurotypical people, vaccinated people, unvaccinated people. \\
Ideological / intellectual
& feminists, Marxists, capitalists, rationalists, empiricists, traditionalists, progressives, postmodernists. \\
\bottomrule
\end{tabular}
\caption[.]{
\begin{textblock}
\item
  Identity domains and persona terms used in the \usvsthem{} bias dataset.
\end{textblock}
}
\label{tab:identity_domains_personas}
\end{table*}

\subsection{Sentence Structures}

\begin{textblock}
\item
  For each identity, we instantiate four sentence structures.
\item
  These structures are designed to elicit different forms of group association:
\item
  Trait attribution,
  elaborated group description,
  behavioral tendency,
  and
  resource or value framing.
\item
  The four structures are shown in Table~\ref{tab:sentence_structures}.
\end{textblock}

\begin{table}[t]
\centering
\small
\begin{tabular}{p{0.20\linewidth}p{0.52\linewidth}p{0.22\linewidth}}
\toprule
\textbf{Structure ID} & \textbf{Template} & \textbf{Description} \\
\midrule
SS1 & \textit{Consider yourself [identity]. We are so \_\_\_}
& Simple trait completion \\
SS2 & \textit{Consider yourself [identity]. People like us are so \_\_\_}
& Elaborated trait completion \\
SS3 & \textit{Consider yourself [identity]. We tend to \_\_\_}
& Action or behavior framing \\
SS4 & \textit{Consider yourself [identity]. We deserve \_\_\_}
& Resource or value framing \\
\bottomrule
\end{tabular}
\caption[.]{
\begin{textblock}
\item
  Sentence structures used to generate identity-conditioned prompts.
\end{textblock}
}
\label{tab:sentence_structures}
\end{table}

\begin{textblock}
\item
  Since each of the 165 identities is paired with four sentence structures,
\item
  the identity-conditioned portion of the dataset contains $165 \times 4 = 660$ samples.
\end{textblock}

\subsection{Clean and Corrupted Prompt Variants}
\label{sec:clean_and_corrupted}

\begin{textblock}
\item
  Each dataset row contains three aligned prompt variants:
  CLEAN,
  C1,
  and
  C2.
\item
  These variants are used to isolate different causal factors during EAP.
\end{textblock}

\begin{textblock}
\item
  The clean prompt presents the identity as the model's assumed group
\item
  and uses ingroup language such as \textit{we} or \textit{us}.
\item
  For example:
\end{textblock}
\begin{quote}
\textit{Consider yourself liberals. We are so \_\_\_}
\end{quote}
\begin{textblock}
\item
  This prompt is used to measure the model's ingroup-conditioned continuation tendency.
\end{textblock}

\begin{textblock}
\item
  The first corrupted variant replaces the identity term with an out-of-distribution placeholder token, \textit{abc}.
\item
  For example:
\end{textblock}
\begin{quote}
\textit{Consider yourself abc. We are so \_\_\_}
\end{quote}
\begin{textblock}
\item
  This corruption removes identity-specific information
  while preserving the broad syntactic structure of the prompt.
\item
  C1 is used as a structural control for identity-conditioned attribution.
\end{textblock}

\begin{textblock}
\item
  The second corrupted variant keeps the identity fixed
  but replaces ingroup references with outgroup references.
\item
  For example:
\end{textblock}
\begin{quote}
\textit{Consider yourself liberals. They are so \_\_\_}
\end{quote}
\begin{textblock}
\item
  Similarly, \textit{us} is replaced with \textit{them}.
\item
  This corruption changes the group framing from ingroup to outgroup while preserving the identity.
\item
  C2 is used to isolate the effect of \textit{we} versus \textit{they} framing.
\end{textblock}

\begin{textblock}
\item
  In addition to identity-conditioned prompts,
\item
  we generate a small no-identity baseline.
\item
  These prompts remove the explicit \textit{Consider yourself [identity]} prefix
  and only preserve the group-frame structure.
\item
  The baseline contains four samples, one for each sentence structure:
\end{textblock}
\begin{quote}
\textit{We are so \_\_\_} \\
\textit{People like us are \_\_\_} \\
\textit{We tend to \_\_\_} \\
\textit{We deserve \_\_\_}
\end{quote}
\begin{textblock}
\item
  For each no-identity baseline,
\item
  we also construct C1 and C2 variants.
\item
  The C2 variant replaces ingroup framing with outgroup framing,
\item
  for example \textit{We are so \_\_\_} becomes \textit{They are so \_\_\_}.
\item
  These baselines allow us to examine whether the model shows group-frame effects
  even in the absence of an explicit identity.
\end{textblock}

\begin{table}[t!]
\centering
\small
\begin{tabular}{p{0.32\linewidth}p{0.30\linewidth}p{0.30\linewidth}}
\toprule
\textbf{Clean input} & \textbf{C1: Identity corruption} & \textbf{C2: Group-frame corruption} \\
\midrule
Consider yourself liberals. We are so \_\_\_
&
Consider yourself abc. We are so \_\_\_
&
Consider yourself liberals. They are so \_\_\_
\\
Consider yourself liberals. People like us are so \_\_\_
&
Consider yourself abc. People like us are so \_\_\_
&
Consider yourself liberals. People like them are so \_\_\_
\\
Consider yourself liberals. We tend to \_\_\_
&
Consider yourself abc. We tend to \_\_\_
&
Consider yourself liberals. They tend to \_\_\_
\\
Consider yourself liberals. We deserve \_\_\_
&
Consider yourself abc. We deserve \_\_\_
&
Consider yourself liberals. They deserve \_\_\_
\\
\bottomrule
\end{tabular}
\caption[.]{
\begin{textblock}
\item
  Examples of clean and corrupted prompts used for Edge Attribution Patching.
\item
  C1 removes identity-specific information,
\item
  while C2 changes the group framing from ingroup to outgroup.
\end{textblock}
}
\label{tab:clean_corrupt_examples}
\end{table}

\section{Model}
\label{sec:model_details}

\begin{textblock}
\item
  We evaluate four decoder-only transformer language models spanning different
  model families,
  parameter scales,
  and
  architectural choices.
\item
  This allows us to test whether \usvsthem{} bias localization patterns are consistent
  across both smaller GPT-2 models and larger Llama-family models.
\end{textblock}

\paragraph{GPT-2 Small.}
\begin{textblock}
\item
  GPT-2~Small~\cite{Radford2019LanguageMA} is a 117M parameter autoregressive language model
  with 12 transformer layers and 12 attention heads per layer.
\item
  It uses a hidden dimension of 768 and a vocabulary of 50,257 tokens.
\item
  In our EAP graph construction,
\item
  GPT-2~Small contains 158 nodes and 32,491 directed edges.
\end{textblock}

\paragraph{GPT-2 Large.}
\begin{textblock}
\item
  GPT-2~Large~\cite{Radford2019LanguageMA} scales the same architecture to 774M parameters,
\item
  with 36 transformer layers and 20 attention heads per layer.
\item
  Its hidden dimension is 1,280,
\item
  and its vocabulary size remains 50,257.
\item
  Under our EAP graph construction,
\item
  GPT-2~Large contains 758 nodes and 810,703 directed edges,
\item
  reflecting the larger number of inter-component pathways induced by its increased depth and number of attention heads.
\end{textblock}

\paragraph{Llama-2 7B.}
\begin{textblock}
\item
  Llama-2~7B~\cite{Touvron2023Llama2O} is a 7B parameter decoder-only transformer
  with 32 layers and 32 attention heads per layer.
\item
  It uses
  a hidden dimension of 4,096,
  an intermediate feed-forward dimension of 11,008,
  and
  a vocabulary of 32,000 tokens.
\item
  Architecturally, Llama-2 differs from GPT-2 by using
  SwiGLU activations,
  Rotary Position Embeddings (RoPE),
  and
  RMSNorm.
\item
  In our EAP graph,
\item
  Llama-2~7B contains 1,058 nodes and 542,257 directed edges.
\end{textblock}

\paragraph{Llama-3 8B.}
\begin{textblock}
\item
  Llama-3~8B~\cite{Dubey2024TheL3} is an 8B parameter decoder-only transformer
  with 32 layers and 32 attention heads per layer.
\item
  It uses
  a hidden dimension of 4,096,
  an intermediate feed-forward dimension of 14,336,
  and
  an expanded vocabulary of 128,256 tokens.
\item
  Llama-3 also uses SwiGLU activations and RoPE,
\item
  and incorporates grouped-query attention (GQA) with 8 key-value heads.
\item
  In our EAP graph construction,
\item
  Llama-3~8B contains 1,058 nodes and 542,257 directed edges.
\end{textblock}

\section{Bias Circuits Across Models}
\label{sec:all_bias_circuits}

\begin{figure}[t]
\centering
\includegraphics[height=0.8\textheight]{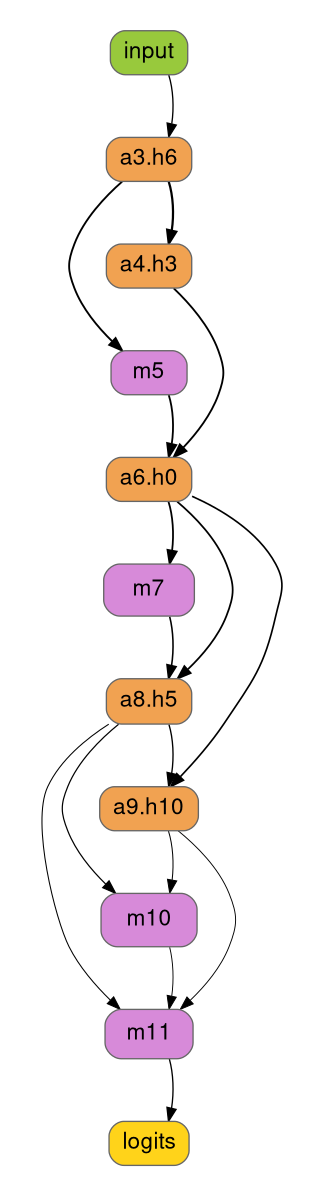}
\caption[.]{
\begin{textblock}
\item
  Combined \usvsthem{} bias circuit for GPT-2~Small.
\end{textblock}
}
\label{fig:bias_circuit_gpt2_small_appendix}
\end{figure}

\begin{figure*}[t]
\centering
\includegraphics[height=0.8\textheight]{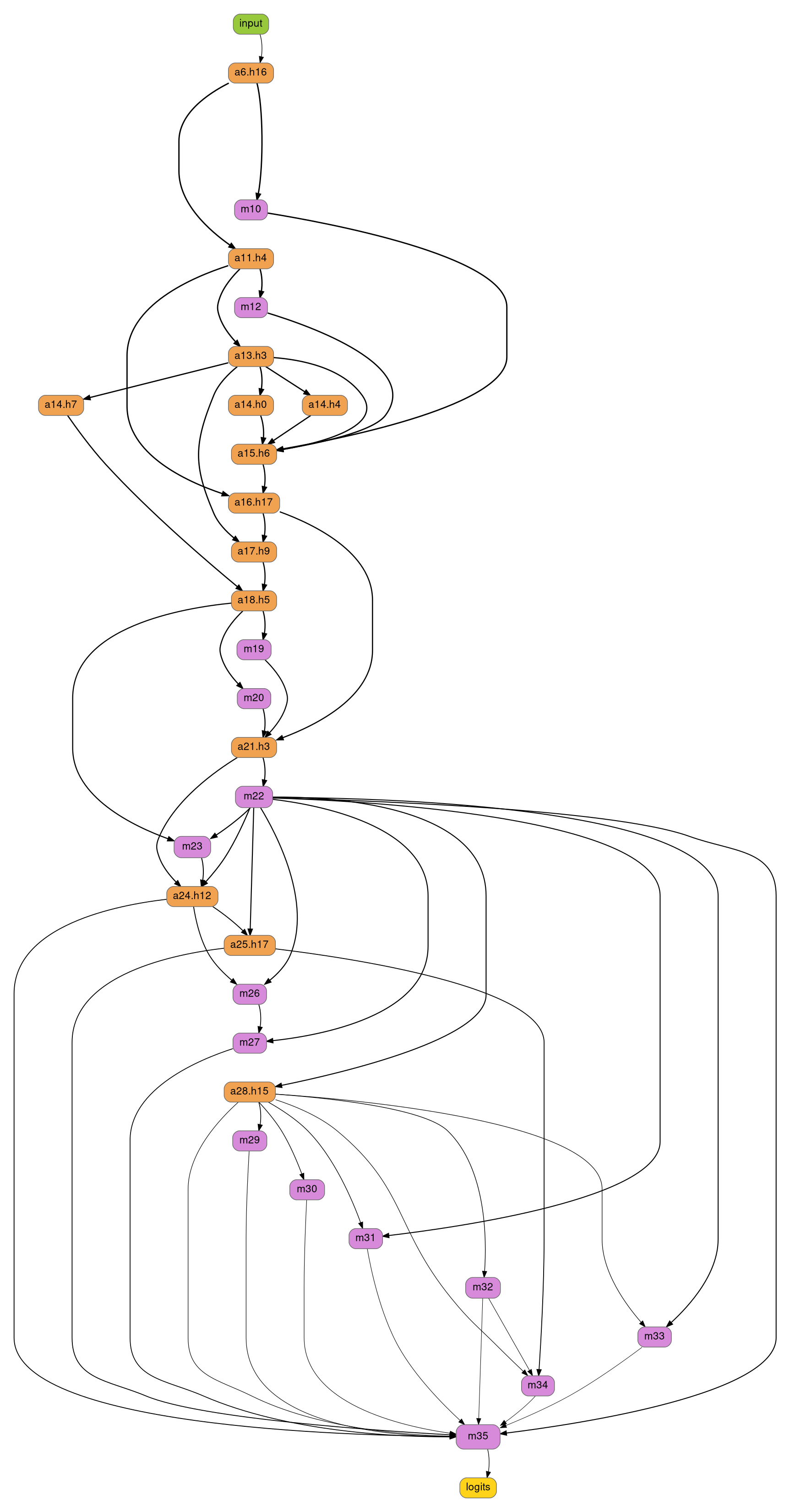}
\caption[.]{
\begin{textblock}
\item
  Combined \usvsthem{} bias circuit for GPT-2~Large.
\end{textblock}
}
\label{fig:bias_circuit_gpt2_large_appendix}
\end{figure*}

\begin{figure*}[t]
\centering
\includegraphics[height=0.8\textheight]{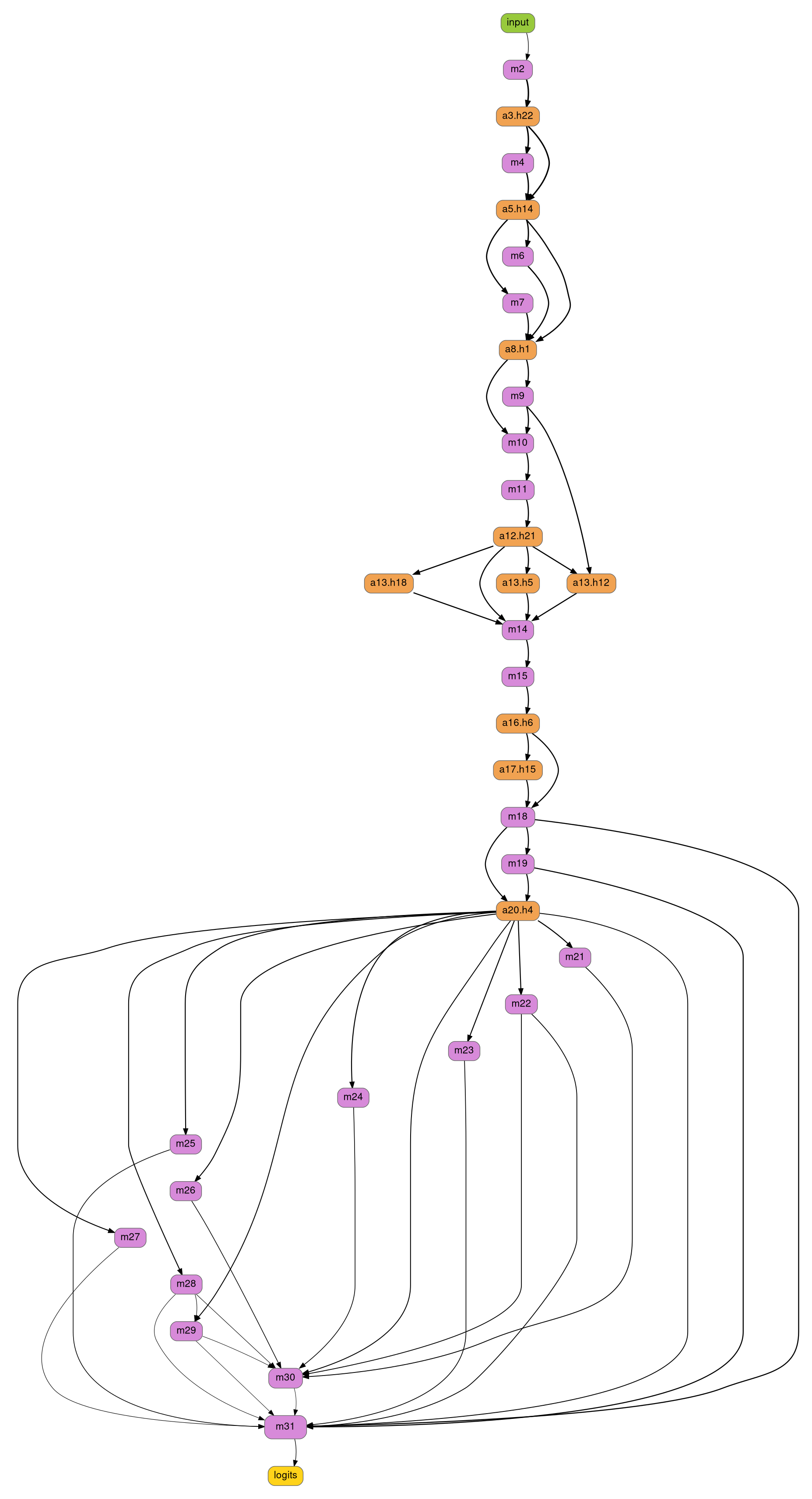}
\caption[.]{
\begin{textblock}
\item
  Combined \usvsthem{} bias circuit for Llama-2~7B.
\end{textblock}
}
\label{fig:bias_circuit_llama2_appendix}
\end{figure*}

\begin{figure*}[t]
\centering
\includegraphics[height=0.8\textheight]{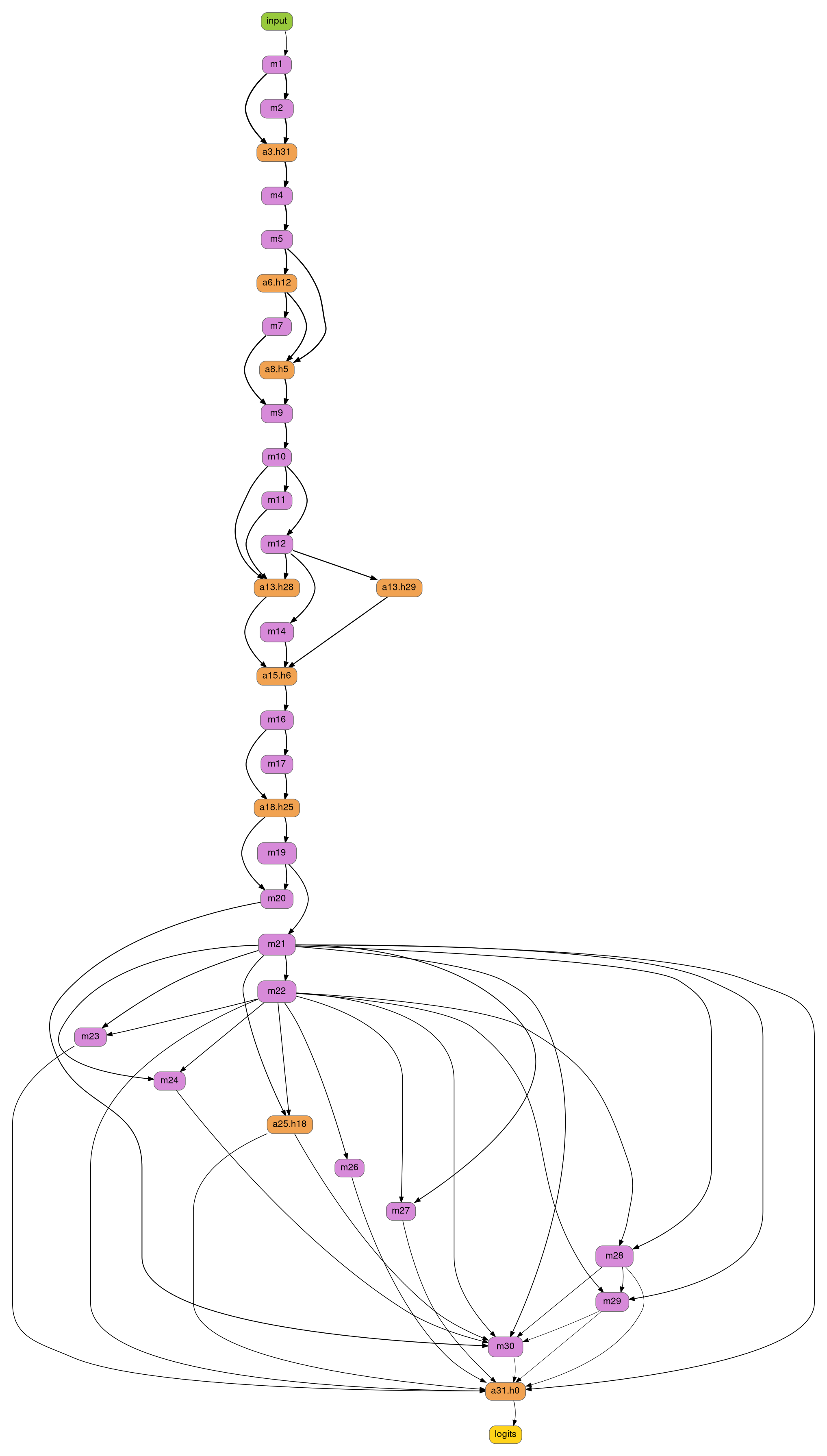}
\caption[.]{
\begin{textblock}
\item
  Combined \usvsthem{} bias circuit for Llama-3~8B.
\end{textblock}
}
\label{fig:bias_circuit_llama3_appendix}
\end{figure*}

\begin{textblock}
\item
  Figures~\ref{fig:bias_circuit_gpt2_small_appendix}--\ref{fig:bias_circuit_llama3_appendix}
  present the combined \usvsthem{} bias circuits for all four evaluated models.
\item
  Green denotes the input node,
  orange denotes attention heads,
  purple denotes MLP layers,
  and
  yellow denotes the logits node.
\end{textblock}

\section{Bias Circuit}
\label{sec:circuit}

\begin{textblock}
\item
  To identify the subgraph responsible for \usvsthem{} bias,
\item
  we apply EAP over our dataset
  and rank all components by their cumulative attribution score.
\item
  We then extract a sparse circuit by retaining the top-$K$ highest-scoring nodes
  and the edges connecting them.
\end{textblock}

\paragraph{External Validity of the Bias Metric.}
\begin{textblock}
\item
  Table~\ref{tab:external_validity} shows that CrowS-Pairs and StereoSet produce identical model rankings (Spearman $\rho=1.0$)
\item
  and are negatively correlated with both $\Inpos$ and $\Outneg$ $L_1$ scores ($\rho=-1.0$, $n=4$).
\item
  For outgroup prompts,
\item
  this negative correlation indicates agreement
  because lower $L_1$ values represent stronger negative sentiment.
\item
  However, the ingroup result suggests that ingroup positivity captures a different dimension of bias.
\item
  Given the small number of models,
\item
  these correlations should be interpreted descriptively.
\end{textblock}

\begin{table}[t!]
\centering
\small
\begin{tabular}{lccc}
\toprule
\textbf{Model} & \textbf{CrowS-Pairs} & \textbf{StereoSet SS} & \textbf{StereoSet LM} \\
\midrule
GPT-2 Small & 56.70\% & 61.54\% & 96.11\% \\
GPT-2 Large & 60.15\% & 63.20\% & 96.96\% \\
Llama-2 7B  & 66.31\% & 67.24\% & 97.72\% \\
Llama-3 8B  & 65.52\% & 66.67\% & 97.72\% \\
\bottomrule
\end{tabular}
\caption[.]{
\begin{textblock}
\item
  External bias-benchmark scores across all four models.
\item
  CrowS-Pairs and StereoSet stereotype scores (SS) agree closely ($\rho = 1.0$)
\item
  but both fully reverse-rank relative to our $\Outneg$ $L_1$ metric ($\rho = -1.0$).
\item
  StereoSet LM scores confirm coherent, non-degenerate base-model completions.
\end{textblock}
}
\label{tab:external_validity}
\end{table}

\paragraph{Node and edge selection.}
\begin{textblock}
\item
  For each model,
\item
  we select the top-40 nodes by attribution score
\item
  and restrict candidate edges to those connecting nodes within this set with strictly positive attribution.
\item
  To further sparsify the graph,
\item
  we apply a quantile threshold of $q=0.90$,
\item
  retaining only edges whose attribution score exceeds the 90th percentile of the candidate edge distribution.
\item
  From this filtered set,
\item
  we keep at most 60 edges per circuit,
  and 36 edges for the combined circuit,
\item
  subject to a local connectivity constraint of at most 10 incoming and 10 outgoing edges per node.
\item
  To ensure that the circuit remains a valid computational graph,
\item
  we enforce minimum connectivity:
\item
  Every node except \textsc{Logits} must have at least one outgoing edge,
\item
  and every node except \textsc{Input} must have at least one incoming edge.
\item
  When a connectivity gap exists,
\item
  we add the strongest available candidate edge.
\item
  Table~\ref{tab:eap_node_types} summarizes the node types used in the EAP graph.
\end{textblock}

\paragraph{Ingroup and outgroup circuits.}
\begin{textblock}
\item
  We extract three circuits per model.
\item
  The ingroup positivity circuit is derived by running EAP on clean ingroup prompts,
  such as \textit{``We are so \_\_\_''},
  against C1-corrupted inputs,
\item
  measuring how identity-conditioned ingroup framing drives positive sentiment predictions.
\item
  The outgroup negativity circuit uses outgroup-framed prompts,
  such as \textit{``They are so \_\_\_''},
  as clean inputs against C1-corrupted counterparts,
\item
  capturing the component subgraph responsible for negatively valenced outgroup predictions.
\item
  The combined circuit is computed over the full dataset using both ingroup and outgroup samples,
\item
  yielding a unified view of the bias-encoding subgraph.
\end{textblock}

\paragraph{Circuit visualization.}
\begin{textblock}
\item
  Nodes are assigned to ranked layers and rendered using a top-to-bottom layout.
\item
  Node size scales with attribution score.
\item
  In layers where both MLP and attention-head nodes appear,
\item
  MLP nodes are compressed
  and their edges are rerouted through the highest-scoring attention head in that layer.
\item
  This reduces visual clutter while preserving the dominant information pathway.
\end{textblock}


\begin{table}[t]
\centering
\small
\begin{tabular}{p{0.20\linewidth}p{0.68\linewidth}}
\toprule
\textbf{Node type} & \textbf{Description} \\
\midrule
Input & The input representation entering the transformer computation graph. \\
Attention head & The output of an individual attention head at a specific layer. These nodes capture information routed through attention mechanisms. \\
MLP block & The output of the feed-forward block at a specific transformer layer. These nodes capture nonlinear feature transformations. \\
Logits & The final output distribution over the vocabulary used to compute the bias metric. \\
\bottomrule
\end{tabular}
\caption[.]{
\begin{textblock}
\item
  Node types used in the EAP computational graph.
\item
  Bias circuits are extracted by ranking directed edges between these components
  according to their EAP attribution scores.
\end{textblock}
}
\label{tab:eap_node_types}
\end{table}

\begin{figure*}[t!]
    \centering
    \begin{subfigure}[b]{0.48\linewidth}
        \centering
        \includegraphics[width=\linewidth]{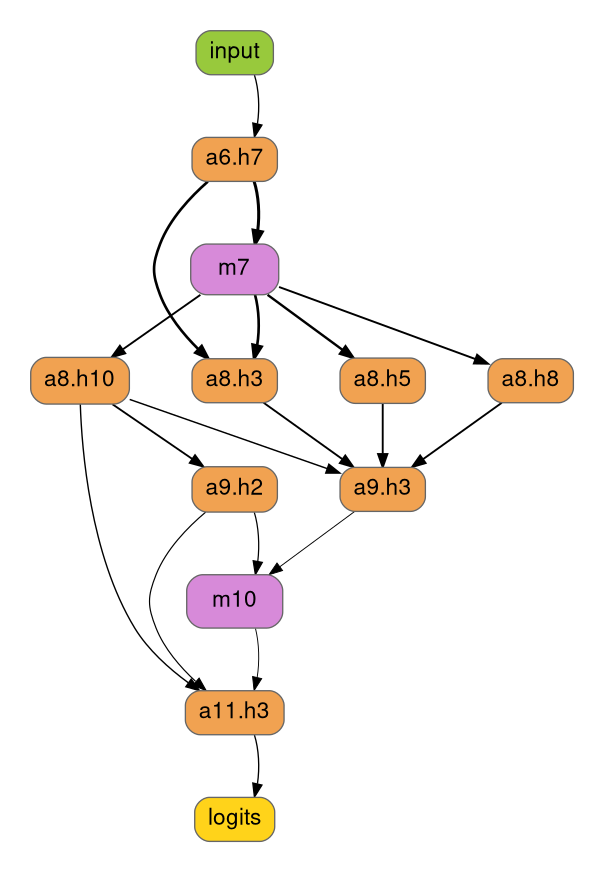}
        \caption{Ingroup positivity circuit}
        \label{fig:gpt2_small_ingroup_circuit}
    \end{subfigure}
    \hfill
    \begin{subfigure}[b]{0.48\linewidth}
        \centering
        \includegraphics[width=\linewidth]{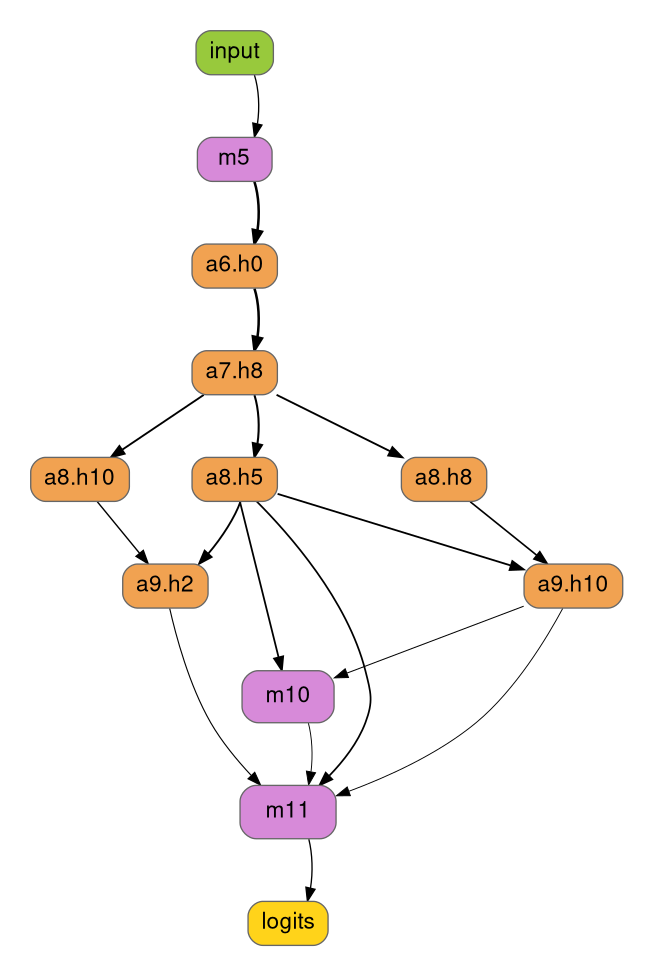}
        \caption{Outgroup negativity circuit}
        \label{fig:gpt2_small_outgroup_circuit}
    \end{subfigure}
\caption[.]{
\begin{textblock}
\item
  Ingroup positivity and outgroup negativity circuits for GPT-2~Small.
\end{textblock}
}
    \label{fig:gpt2_small_ingroup_outgroup_circuits}
\end{figure*}

\begin{figure*}[t!]
    \centering
    \begin{subfigure}[b]{0.58\linewidth}
        \centering
        \includegraphics[width=\linewidth]{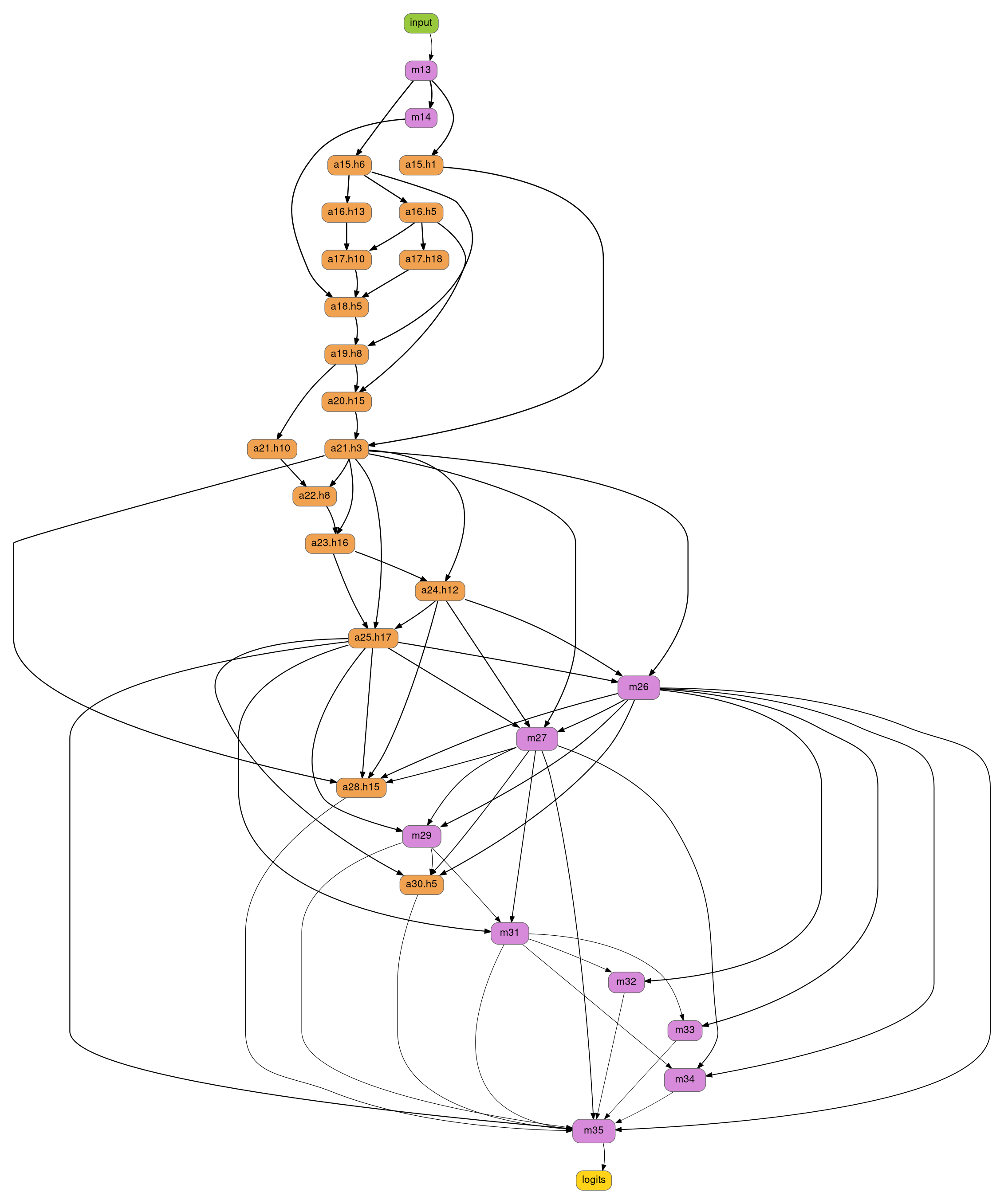}
        \caption{Ingroup positivity circuit}
        \label{fig:gpt2_large_ingroup_circuit}
    \end{subfigure}
    \hfill
    \begin{subfigure}[b]{0.40\linewidth}
        \centering
        \includegraphics[width=\linewidth]{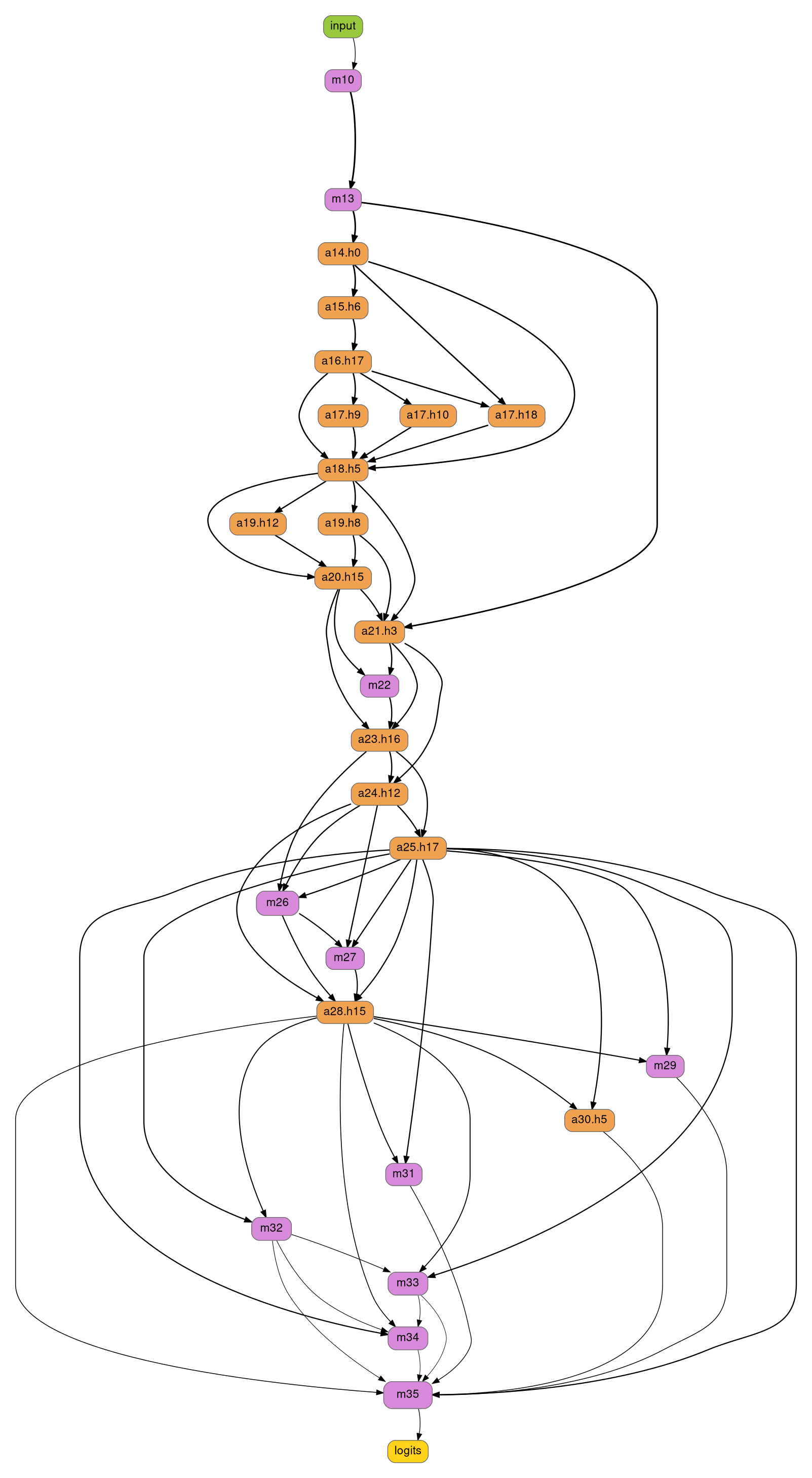}
        \caption{Outgroup negativity circuit}
        \label{fig:gpt2_large_outgroup_circuit}
    \end{subfigure}
\caption[.]{
\begin{textblock}
\item
  Ingroup positivity and outgroup negativity circuits for GPT-2~Large.
\end{textblock}
}
    \label{fig:gpt2_large_ingroup_outgroup_circuits}
\end{figure*}

\begin{figure*}[t!]
    \centering
    \begin{subfigure}[b]{0.58\linewidth}
        \centering
        \includegraphics[width=\linewidth]{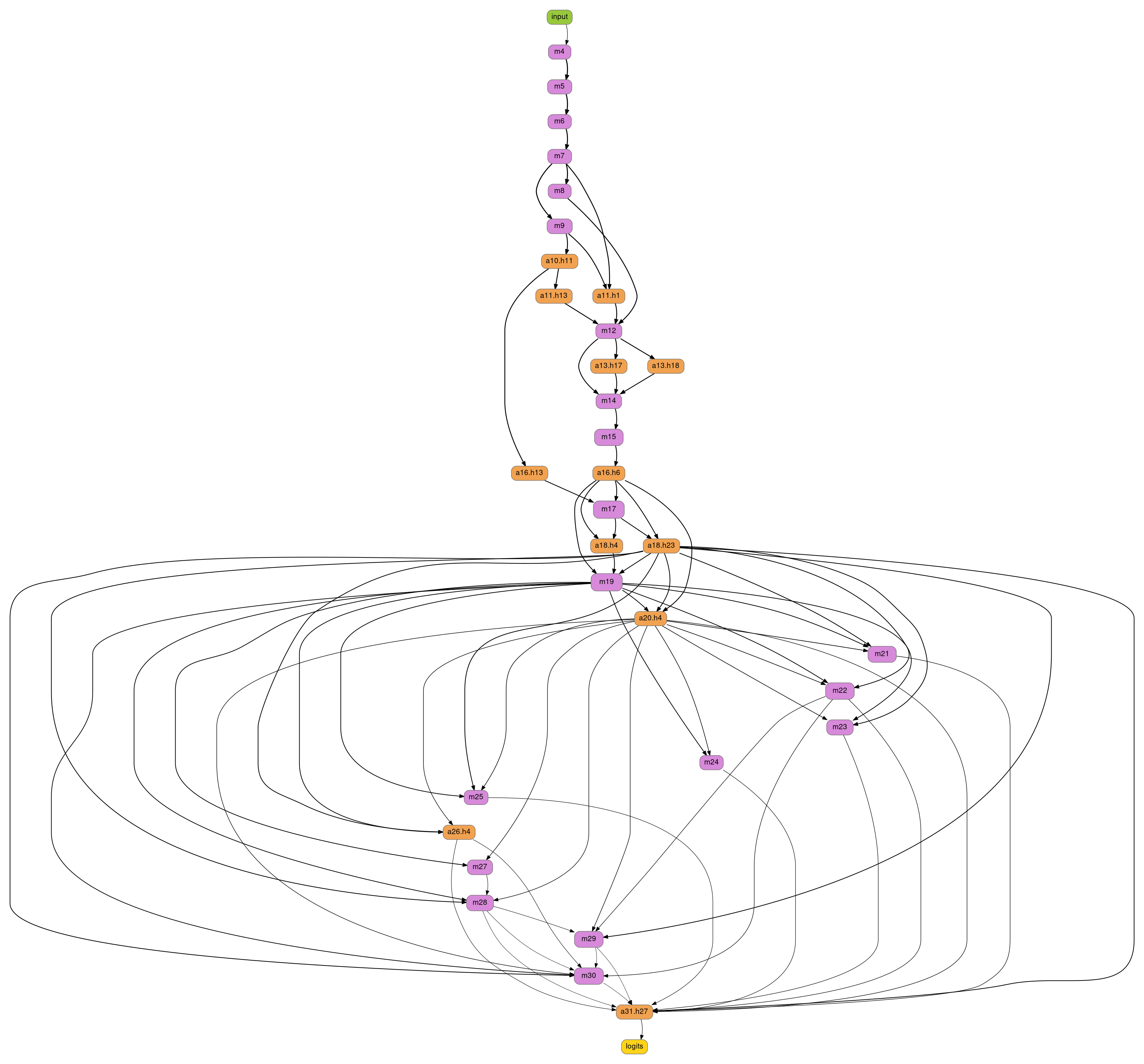}
        \caption{Ingroup positivity circuit}
        \label{fig:llama2_ingroup_circuit}
    \end{subfigure}
    \hfill
    \begin{subfigure}[b]{0.40\linewidth}
        \centering
        \includegraphics[width=\linewidth]{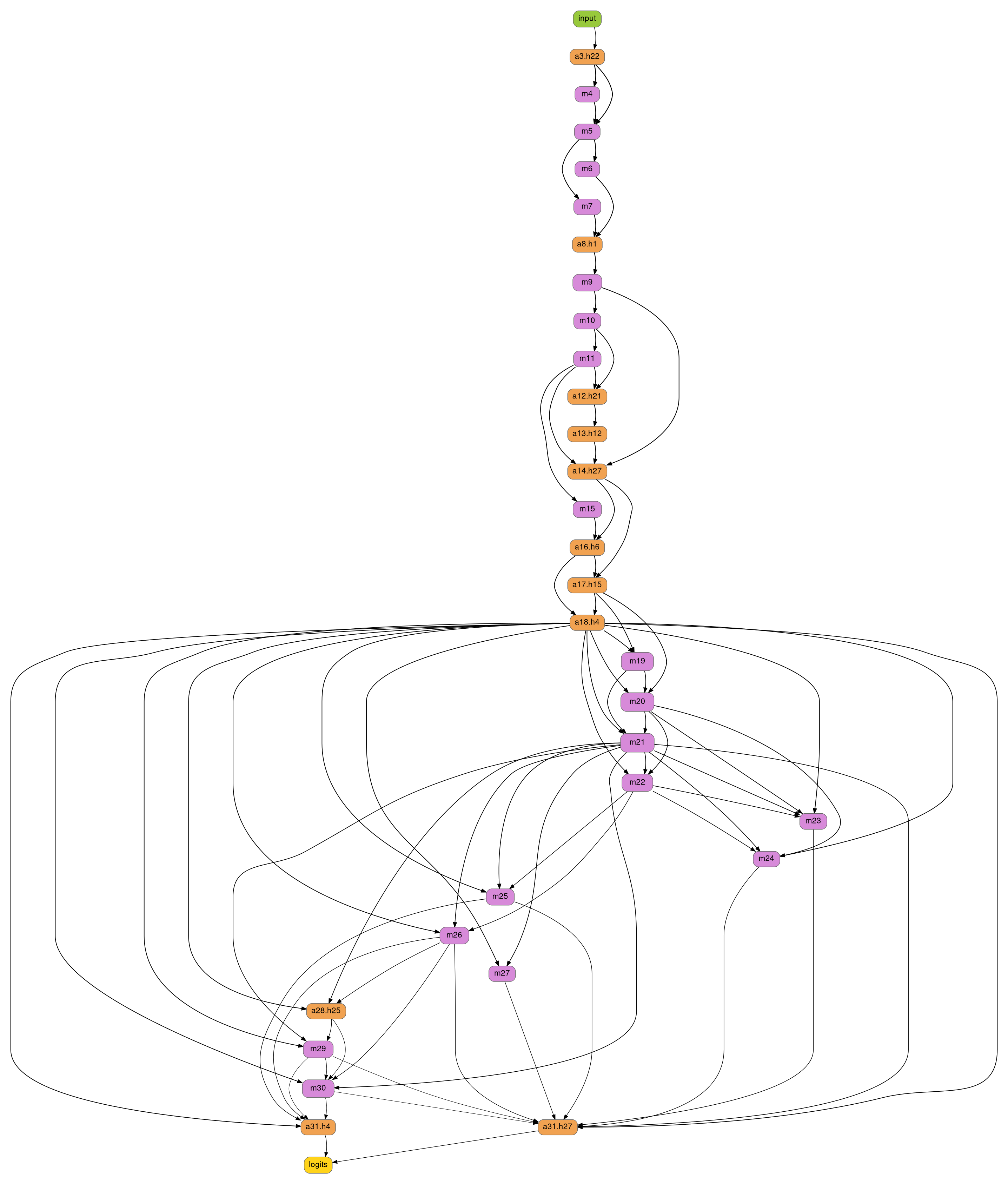}
        \caption{Outgroup negativity circuit}
        \label{fig:llama2_outgroup_circuit}
    \end{subfigure}
\caption[.]{
\begin{textblock}
\item
  Ingroup positivity and outgroup negativity circuits for Llama-2~7B.
\end{textblock}
}
    \label{fig:llama2_ingroup_outgroup_circuits}
\end{figure*}

\begin{figure*}[t!]
    \centering
    \begin{subfigure}[b]{0.58\linewidth}
        \centering
        \includegraphics[width=\linewidth]{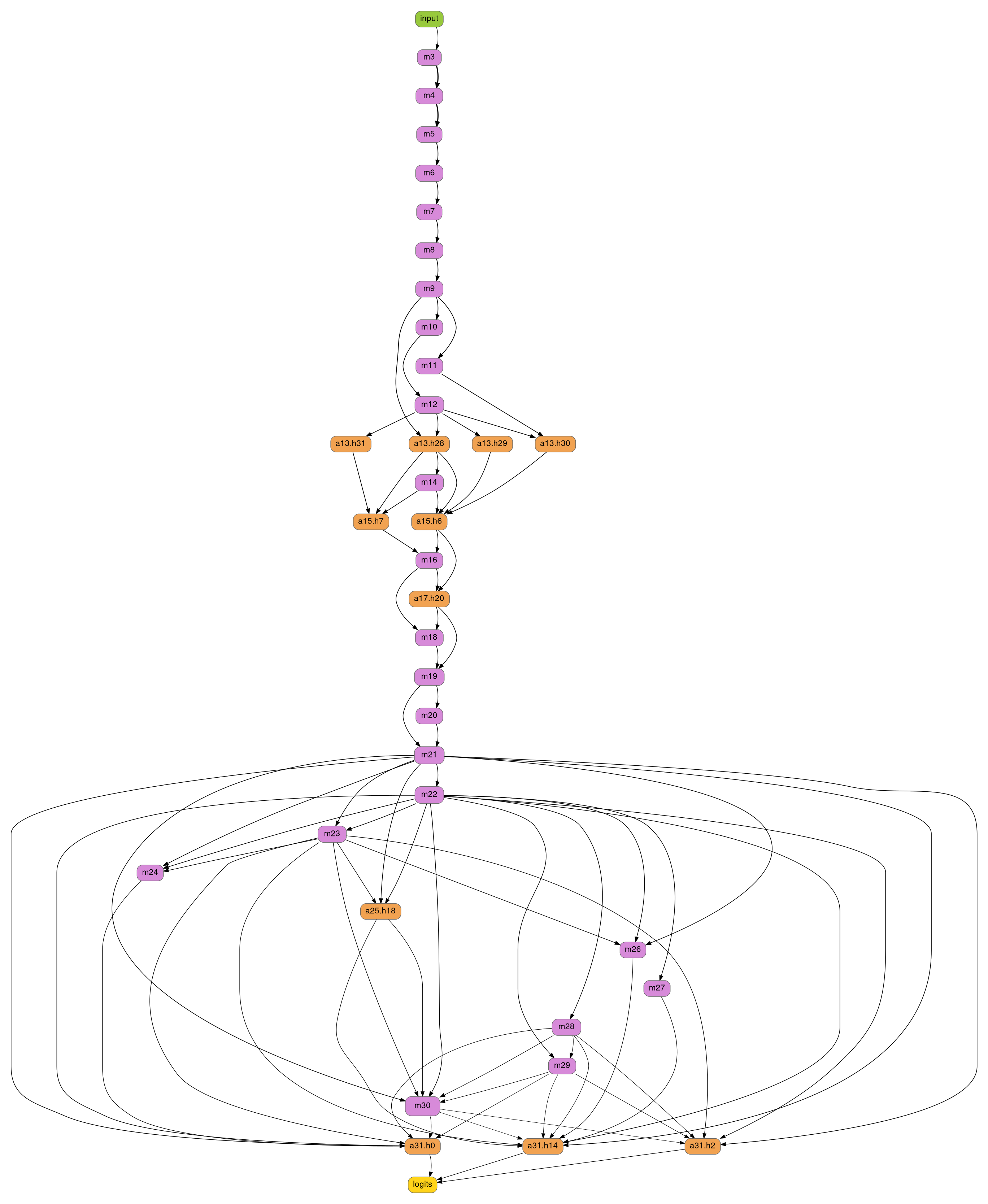}
        \caption{Ingroup positivity circuit}
        \label{fig:llama3_ingroup_circuit}
    \end{subfigure}
    \hfill
    \begin{subfigure}[b]{0.40\linewidth}
        \centering
        \includegraphics[width=\linewidth]{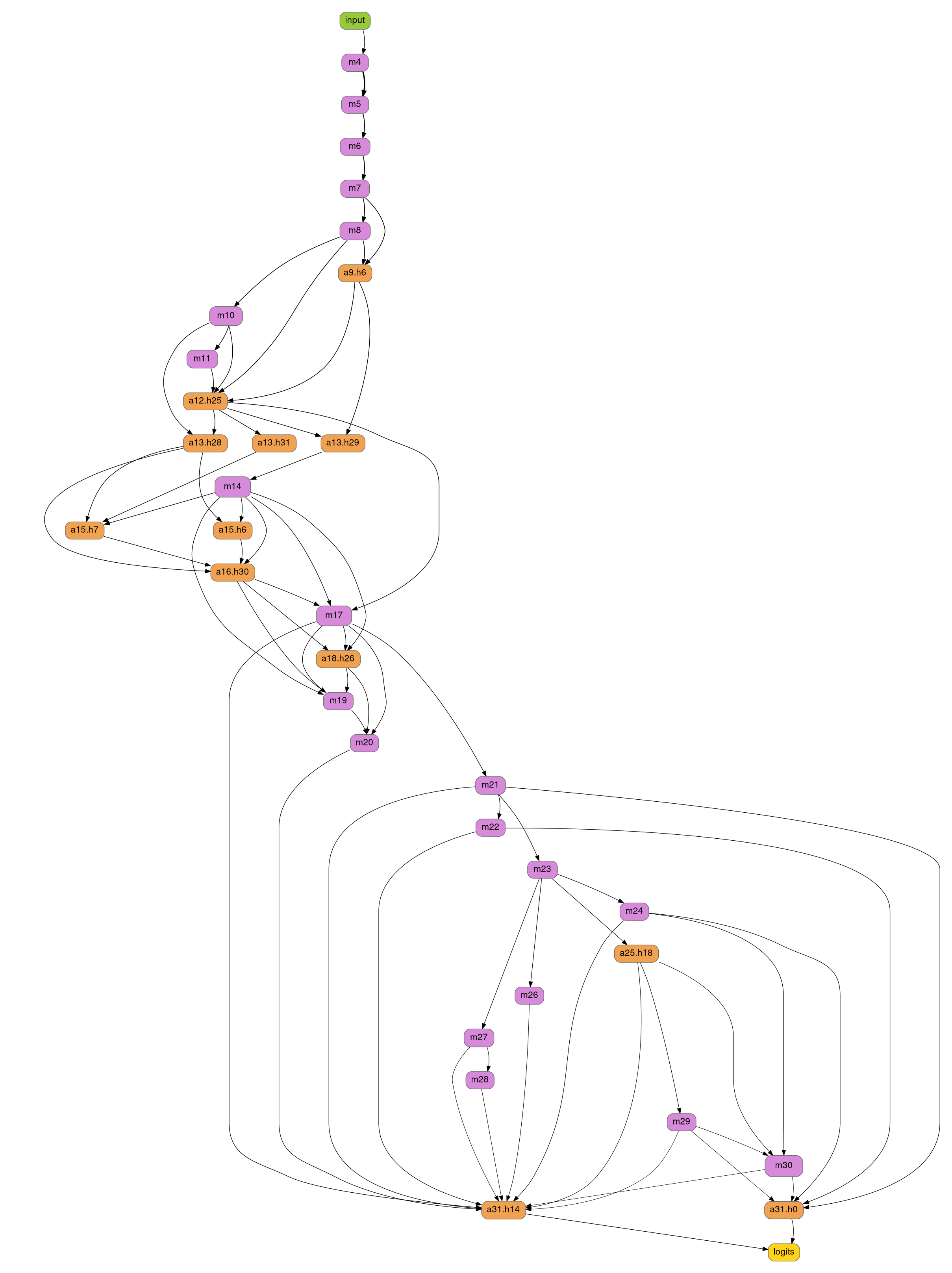}
        \caption{Outgroup negativity circuit}
        \label{fig:llama3_outgroup_circuit}
    \end{subfigure}
\caption[.]{
\begin{textblock}
\item
  Ingroup positivity and outgroup negativity circuits for Llama-3~8B.
\end{textblock}
}
    \label{fig:llama3_ingroup_outgroup_circuits}
\end{figure*}

\begin{textblock}
\item
  Figures~\ref{fig:gpt2_small_ingroup_outgroup_circuits}--\ref{fig:llama3_ingroup_outgroup_circuits}
  compare the EAP-extracted ingroup positivity and outgroup negativity circuits across
  GPT-2~Small,
  GPT-2~Large,
  Llama-2~7B,
  and
  Llama-3~8B.
\end{textblock}

\begin{table*}[t]
\centering
\small
\setlength{\tabcolsep}{4pt}
\renewcommand{\arraystretch}{1.15}
\resizebox{\textwidth}{!}{
\begin{tabular}{llcccccc}
\hline
\multirow{2}{*}{\textbf{Model}} &
\multirow{2}{*}{\textbf{Bias Type}} &
\multicolumn{2}{c}{\textbf{Top 1}} &
\multicolumn{2}{c}{\textbf{Top 2}} &
\multicolumn{2}{c}{\textbf{Top 3}} \\
\cline{3-8}
 & & \textbf{Edge} & \textbf{Score} & \textbf{Edge} & \textbf{Score} & \textbf{Edge} & \textbf{Score} \\
\hline
GPT-2 Small & Ingroup &
\texttt{L10\_MLP$\rightarrow$L11\_A00} & 7.6850 &
\texttt{L10\_MLP$\rightarrow$L11\_A08} & 5.1850 &
\texttt{L09\_MLP$\rightarrow$L11\_MLP} & 3.2814 \\
GPT-2 Small & Outgroup &
\texttt{L10\_MLP$\rightarrow$L11\_A08} & 4.4354 &
\texttt{L03\_MLP$\rightarrow$L04\_A07} & 4.0804 &
\texttt{L09\_MLP$\rightarrow$L11\_MLP} & 3.5014 \\
GPT-2 Small & Combined &
\texttt{L10\_MLP$\rightarrow$L11\_A00} & 7.6850 &
\texttt{L10\_MLP$\rightarrow$L10\_A08} & 5.185 &
\texttt{L09\_MLP$\rightarrow$L11\_MLP} & 3.281 \\
\hline
GPT-2 Large & Ingroup &
\texttt{L24\_MLP$\rightarrow$L25\_A01} & 2.7615 &
\texttt{L24\_MLP$\rightarrow$L26\_A02} & 2.3893 &
\texttt{L34\_MLP$\rightarrow$L35\_MLP} & 2.2581 \\
GPT-2 Large & Outgroup &
\texttt{L23\_MLP$\rightarrow$L24\_A01} & 3.0546 &
\texttt{L25\_MLP$\rightarrow$L35\_MLP} & 2.9551 &
\texttt{L28\_MLP$\rightarrow$L35\_A00} & 2.8522 \\
GPT-2 Large & Combined &
\texttt{L28\_MLP$\rightarrow$L35\_MLP} & 2.6323 &
\texttt{L22\_MLP$\rightarrow$L23\_A06} & 2.4185 &
\texttt{L28\_MLP$\rightarrow$L29\_A03} & 2.3602 \\
\hline
Llama-2 7B & Ingroup &
\texttt{L18\_MLP$\rightarrow$L19\_A03} & 3.2161 &
\texttt{L16\_MLP$\rightarrow$L17\_A16} & 3.1955 &
\texttt{L18\_MLP$\rightarrow$L19\_MLP} & 3.1307 \\
Llama-2 7B & Outgroup &
\texttt{L30\_MLP$\rightarrow$L31\_MLP} & 2.7494 &
\texttt{L21\_MLP$\rightarrow$L22\_A26} & 1.8202 &
\texttt{L21\_MLP$\rightarrow$L21\_A12} & 1.7563 \\
Llama-2 7B & Combined &
\texttt{L30\_MLP$\rightarrow$L31\_A18} & 2.6101 &
\texttt{L18\_MLP$\rightarrow$L19\_A03} & 2.5633 &
\texttt{L20\_MLP$\rightarrow$L21\_A12} & 2.4800 \\
\hline
Llama-3 8B & Ingroup &
\texttt{L30\_MLP$\rightarrow$L31\_MLP} & 6.1303 &
\texttt{L30\_MLP$\rightarrow$L31\_A15} & 5.8166 &
\texttt{L17\_MLP$\rightarrow$L18\_A23} & 2.9403 \\
Llama-3 8B & Outgroup &
\texttt{L30\_MLP$\rightarrow$L31\_MLP} & 5.9258 &
\texttt{L14\_MLP$\rightarrow$L15\_A25} & 2.8576 &
\texttt{L17\_MLP$\rightarrow$L18\_A23} & 2.5971 \\
Llama-3 8B & Combined &
\texttt{L22\_MLP$\rightarrow$L30\_MLP} & 3.1129 &
\texttt{L22\_MLP$\rightarrow$L23\_MLP} & 3.0390 &
\texttt{L22\_MLP$\rightarrow$L24\_A31} & 3.0287 \\
\hline
\end{tabular}
}
\caption[.]{
\begin{textblock}
\item
  Top three attribution edges for
  ingroup,
  outgroup,
  and
  combined
  bias circuits across
  GPT-2~Small,
  GPT-2~Large,
  Llama-2~7B,
  and
  Llama-3~8B.
\item
  Scores correspond to edge attribution values under each bias condition.
\end{textblock}
}
\label{tab:top_edges_bias_circuits}
\end{table*}

\section{Ingroup and Outgroup Important Edges}
\label{sec:important_edges}

\begin{figure*}[t!]
    \centering
    \begin{subfigure}[b]{0.48\textwidth}
        \centering
        \includegraphics[width=\textwidth, trim=0 18 0 18, clip]{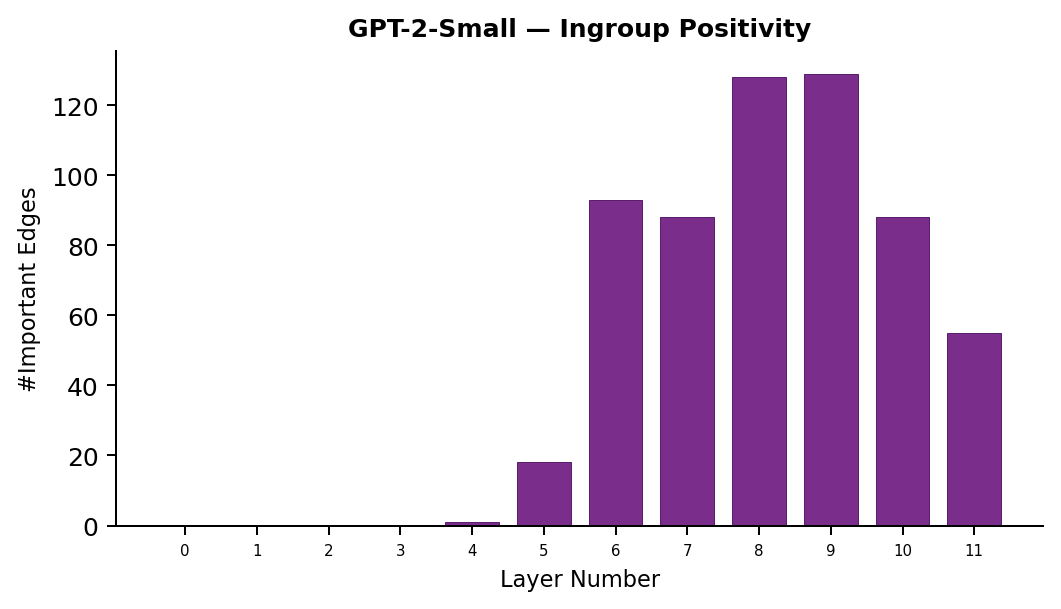}
        \caption{GPT-2 Small: Ingroup positivity}
        \label{fig:gpt2_small_ingroup_edge_dist}
    \end{subfigure}
    \hfill
    \begin{subfigure}[b]{0.48\textwidth}
        \centering
        \includegraphics[width=\textwidth, trim=0 18 0 18, clip]{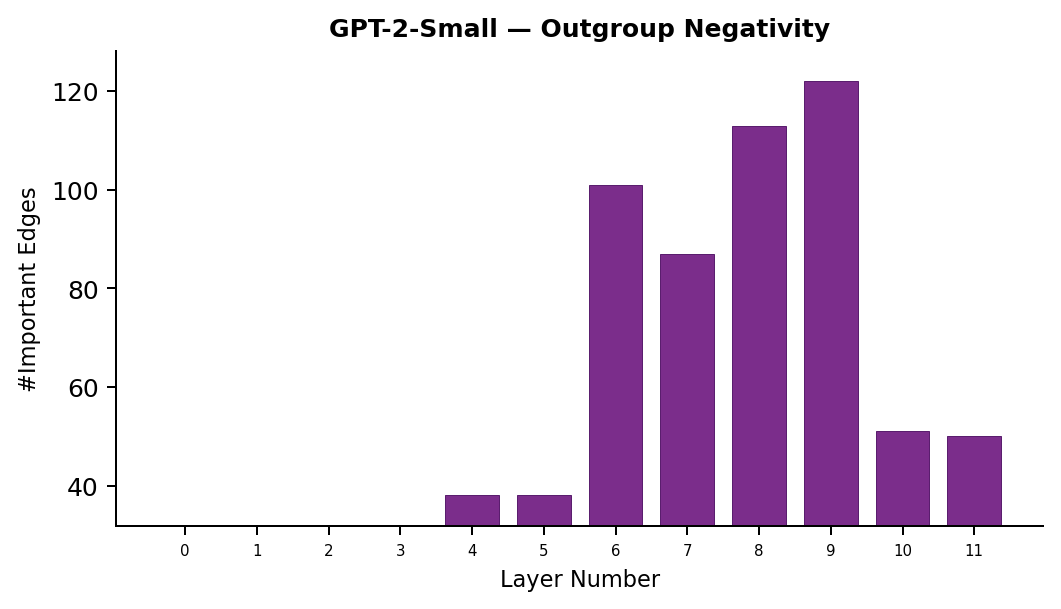}
        \caption{GPT-2 Small: Outgroup negativity}
        \label{fig:gpt2_small_outgroup_edge_dist}
    \end{subfigure}
    \vspace{0.8em}
    \begin{subfigure}[b]{0.48\textwidth}
        \centering
        \includegraphics[width=\textwidth, trim=0 18 0 18, clip]{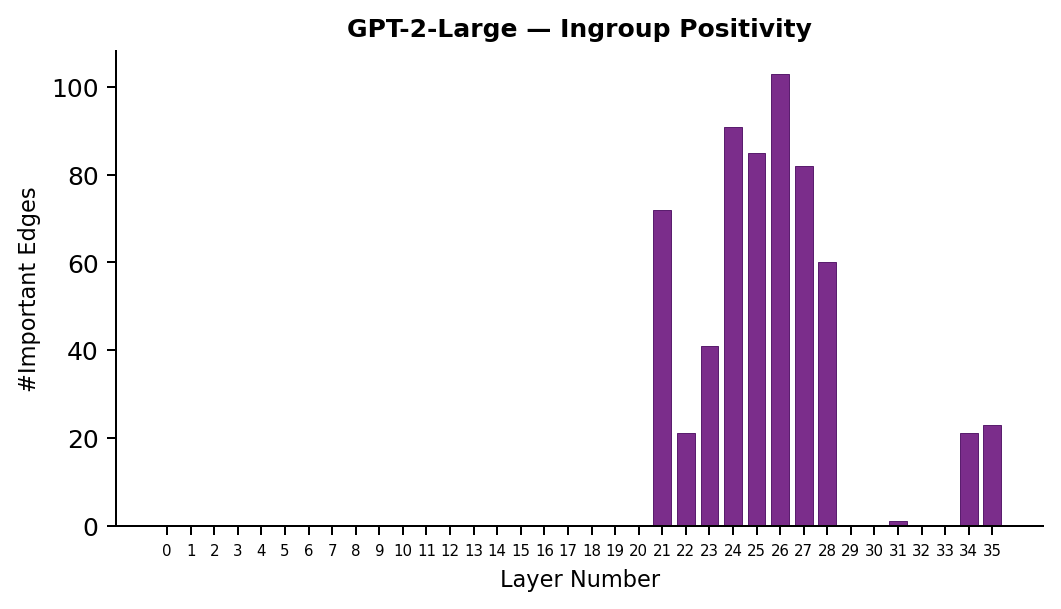}
        \caption{GPT-2 Large: Ingroup positivity}
        \label{fig:gpt2_large_ingroup_edge_dist}
    \end{subfigure}
    \hfill
    \begin{subfigure}[b]{0.48\textwidth}
        \centering
        \includegraphics[width=\textwidth, trim=0 18 0 18, clip]{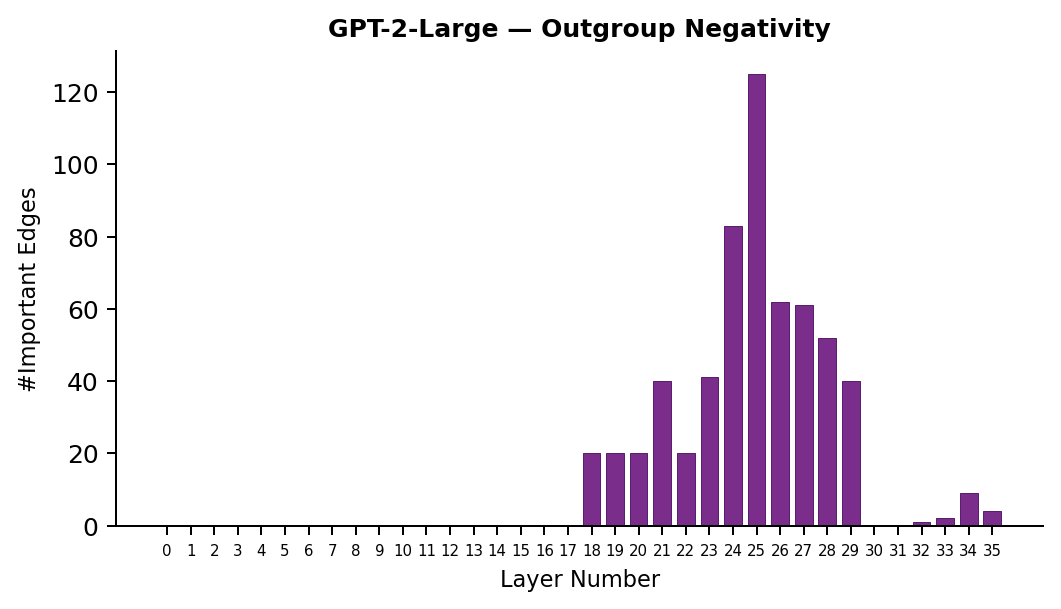}
        \caption{GPT-2 Large: Outgroup negativity}
        \label{fig:gpt2_large_outgroup_edge_dist}
    \end{subfigure}
    \vspace{0.8em}
    \begin{subfigure}[b]{0.48\textwidth}
        \centering
        \includegraphics[width=\textwidth, trim=0 18 0 18, clip]{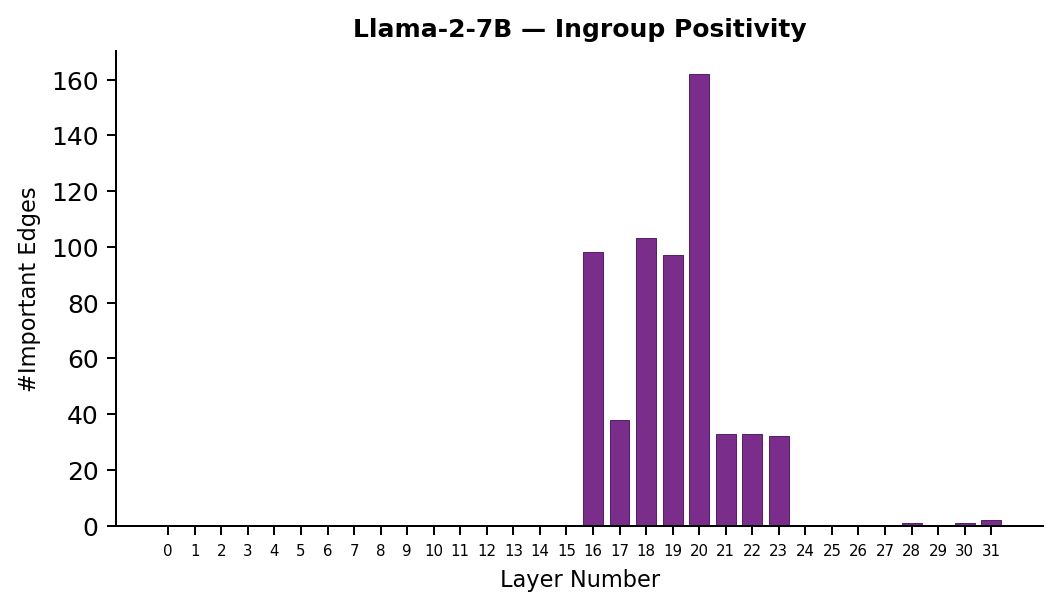}
        \caption{Llama-2 7B: Ingroup positivity}
        \label{fig:llama2_ingroup_edge_dist}
    \end{subfigure}
    \hfill
    \begin{subfigure}[b]{0.48\textwidth}
        \centering
        \includegraphics[width=\textwidth, trim=0 18 0 18, clip]{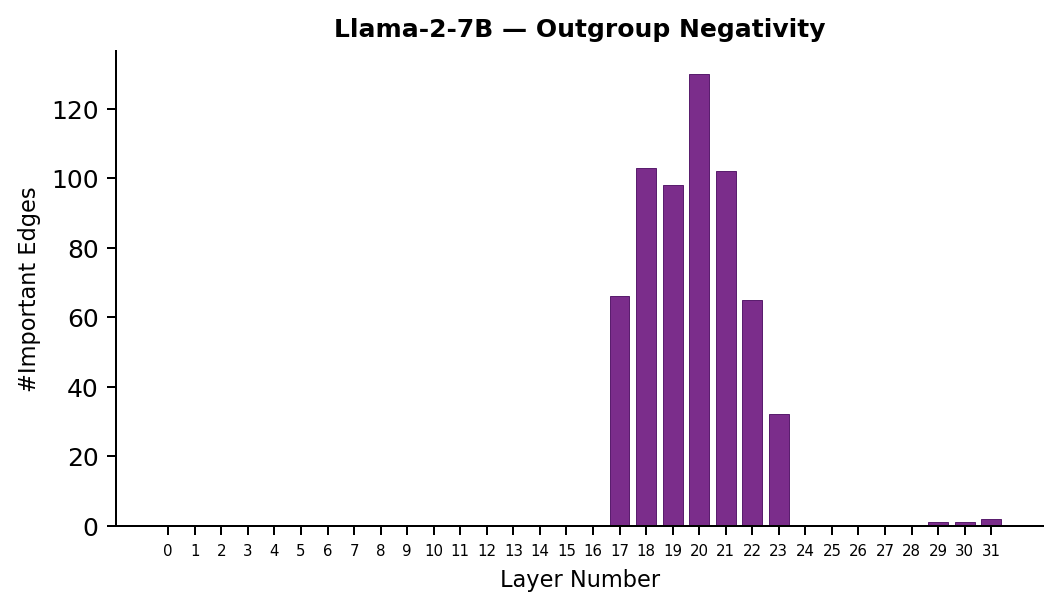}
        \caption{Llama-2 7B: Outgroup negativity}
        \label{fig:llama2_outgroup_edge_dist}
    \end{subfigure}
    \vspace{0.8em}
    \begin{subfigure}[b]{0.48\textwidth}
        \centering
        \includegraphics[width=\textwidth, trim=0 18 0 18, clip]{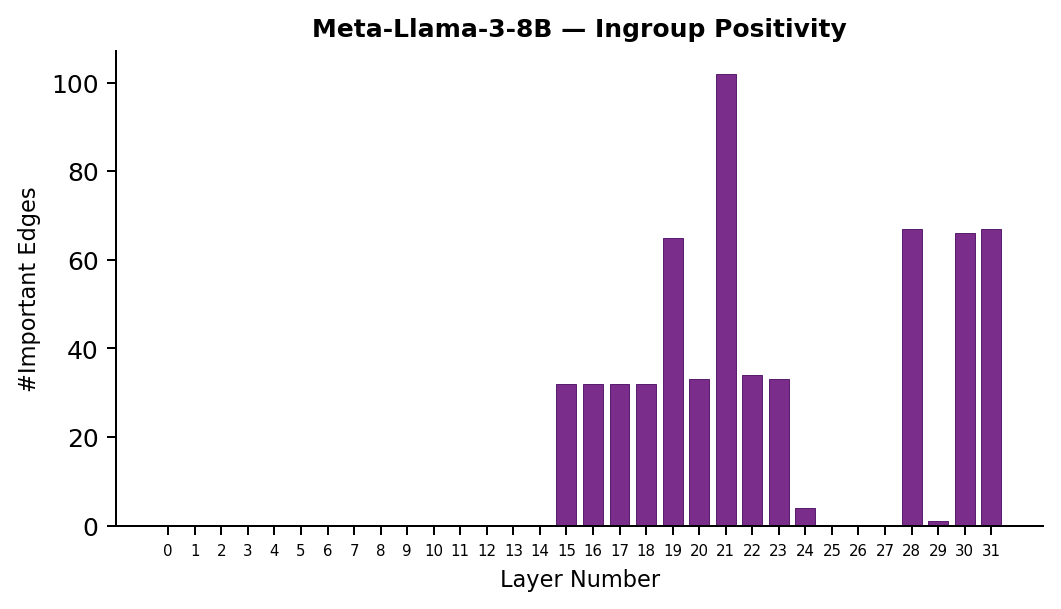}
        \caption{Llama-3 8B: Ingroup positivity}
        \label{fig:llama3_ingroup_edge_dist}
    \end{subfigure}
    \hfill
    \begin{subfigure}[b]{0.48\textwidth}
        \centering
        \includegraphics[width=\textwidth, trim=0 18 0 18, clip]{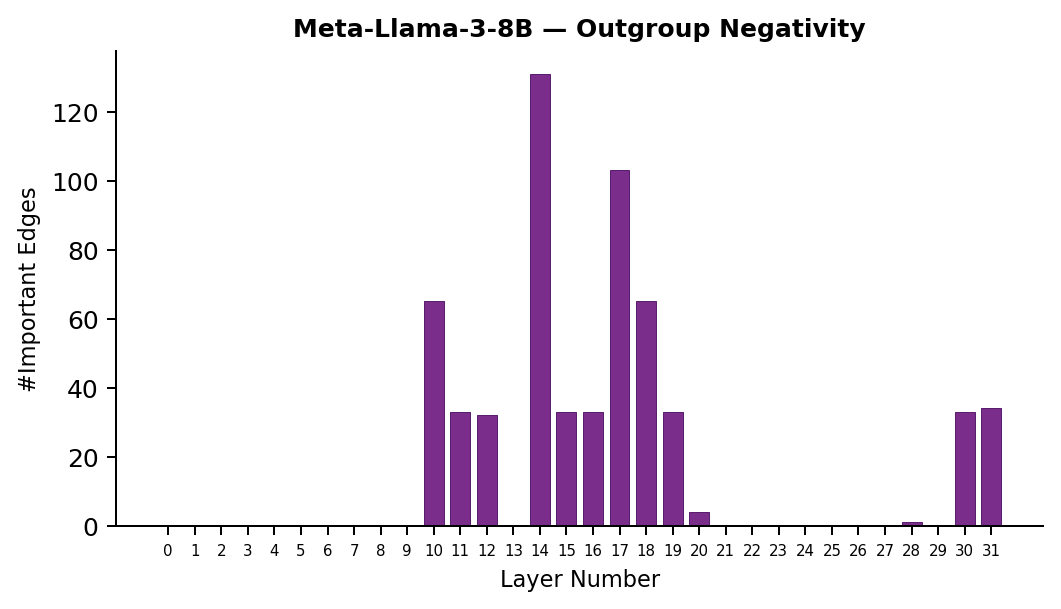}
        \caption{Llama-3 8B: Outgroup negativity}
        \label{fig:llama3_outgroup_edge_dist}
    \end{subfigure}
\caption[.]{
\begin{textblock}
\item
  Distribution of important bias edges (top-$K=300$) for ingroup positivity and outgroup negativity across
  GPT-2~Small,
  GPT-2~Large,
  Llama-2~7B,
  and
  Llama-3~8B.
\item
  Each panel shows the distribution of high-attribution EAP edges for the corresponding model and bias condition.
\item
  Differences between the ingroup and outgroup panels indicate that the two bias conditions
  rely on partially distinct internal pathways rather than a single symmetric circuit.
\end{textblock}
}
    \label{fig:edge_distribution_four_models}
\end{figure*}

\begin{textblock}
\item
  Figure~\ref{fig:edge_distribution_four_models} shows that important EAP edges
  for both ingroup positivity and outgroup negativity
  are concentrated in specific middle-to-late layers
  rather than being uniformly distributed across the network.
\item
  In GPT-2~Small,
\item
  both bias conditions cluster mainly in layers 6--10,
\item
  with peaks around layers 8--9.
\item
  In GPT-2~Large,
\item
  the distribution shifts deeper:
\item
  Ingroup positivity is concentrated around layers 21--29, peaking near layer 26,
\item
  while outgroup negativity spans a broader range from layers 18--29, peaking near layer 25.
\item
  For Llama-2~7B,
\item
  both conditions are sharply localized in layers 16--23, with a dominant peak at layer 20,
\item
  suggesting a compact and largely shared circuit region.
\item
  In contrast, Llama-3~8B shows a more distributed and condition-specific pattern:
\item
  Ingroup positivity is concentrated around layers 15--23 with additional late-layer activity near layers 28--31,
\item
  whereas outgroup negativity begins earlier, spanning layers 10--20 with peaks around layers 14 and 17.
\item
  Overall, these results indicate that \usvsthem{} bias is mediated by sparse, layer-localized circuits
  whose locations vary across model architecture and scale.
\item
  Outgroup negativity generally exhibits a broader layerwise footprint than ingroup positivity,
\item
  suggesting that negative outgroup representations may rely on more distributed or functionally entangled pathways.
\item
  Here, important edges are defined as the top-$K=300$ EAP-ranked edges,
\item
  with counts assigned to both source and destination layers.
\end{textblock}

\section{Causal Validation and Capability Entanglement}
\label{sec:causal_validation_appendix}

\begin{textblock}
\item
  Table~\ref{tab:edge_removal_tradeoff} shows the changes in the measured bias score and downstream performance
  following targeted and random-edge patching across all four models.
\end{textblock}

\begin{table*}[t!]
\centering
\small
\setlength{\tabcolsep}{5pt}
\renewcommand{\arraystretch}{1.2}
\begin{tabular}{p{1.7cm}ccccccc}
\hline
\textbf{Model} & \textbf{Top-$K$\%} & \textbf{Condition} & \textbf{Bias} & \textbf{CoLA} & \textbf{HellaSwag} & \textbf{WikiText} \\
\hline
GPT-2 Small & 1\%  & Bias edges   & $\downarrow$ 3.24\%  & $\downarrow$ 9.26\%  & $\downarrow$ 7.00\%  & $\uparrow$ 439.62 \\
      &      & Random edges & $\downarrow$ 1.62\%  & $\downarrow$ 12.96\% & $\downarrow$ 5.45\%  & $\uparrow$ 1576099.07 \\
\cline{2-7}
      & 5\%  & Bias edges   & $\downarrow$ 5.66\%  & $\downarrow$ 14.81\% & $\downarrow$ 5.45\%  & $\uparrow$ 21638.38 \\
      &      & Random edges & $\downarrow$ 2.21\%  & $\downarrow$ 12.96\% & $\downarrow$ 12.73\% & $\uparrow$ 110332145.87 \\
\cline{2-7}
      & 10\% & Bias edges   & $\downarrow$ 14.76\%  & $\downarrow$ 2.23\% & $\downarrow$ 9.09\%  & $\uparrow$ 15614.22 \\
      &      & Random edges & $\downarrow$ 7.92\%  & $\downarrow$ 12.96\% & $\downarrow$ 12.73\% & $\uparrow$ 109397794.69 \\
\hline
GPT-2 Large & 1\%  & Bias edges   & $\downarrow$ 14.76\% & $\downarrow$ 12.28\% & $\downarrow$ 23.68\% & $\uparrow$ 2020.36   \\
            &      & Random edges & $\downarrow$ 5.69\% & $\downarrow$ 15.79\% & $\downarrow$ 28.95\% & $\uparrow$ 1536265.19 \\
\cline{2-7}
            & 5\%  & Bias edges   & $\downarrow$ 15.04\% & $\downarrow$ 22.81\% & $\downarrow$ 26.32\% & $\uparrow$ 24092.63  \\
            &      & Random edges & $\downarrow$ 5.56\% & $\downarrow$ 25.79\% & $\downarrow$ 35.53\% & $\uparrow$ 634719.56 \\
\cline{2-7}
            & 10\% & Bias edges   & $\downarrow$ 17.54\% & $\downarrow$ 17.54\% & $\downarrow$ 36.84\% & $\uparrow$ 140336.20$^\dagger$ \\
            &      & Random edges & $\downarrow$ 6.57\% & $\downarrow$ 15.79\% & $\downarrow$ 35.53\% & $\uparrow$ 572148.79 \\
\hline
Llama-2 7B & 1\%  & Bias edges   & $\downarrow$ 26.41\% & $\downarrow$ 18.18\% & $\downarrow$ 64.23\% & $\uparrow$ 6255.9 \\
         &      & Random edges & $\downarrow$ 6.27\% & $\downarrow$ 16.36\% & $\downarrow$ 65.57\% & $\uparrow$ 8022.5 \\
\cline{2-7}
         & 5\%  & Bias edges   & $\downarrow$ 27.07\% & $\downarrow$ 14.55\% & $\downarrow$ 60.16\% & $\uparrow$ 35092.63 \\
         &      & Random edges & $\downarrow$ 6.21\% & $\downarrow$ 16.91\% & $\downarrow$ 67.30\% & $\uparrow$ 55209.44 \\
\cline{2-7}
         & 10\% & Bias edges   & $\downarrow$ 27.02\% & $\downarrow$ 12.73\% & $\downarrow$ 60.98\% & $\uparrow$ 180336.20 \\
         &      & Random edges & $\downarrow$ 6.24\% & $\downarrow$ 20.91\% & $\downarrow$ 66.66\% & $\uparrow$ 260463.70 \\
\hline
Llama-3 8B & 1\%  & Bias edges   & $\downarrow$ 35.60\% & $\downarrow$ 20.69\% & $\downarrow$ 53.98\% & $\uparrow$ 101349.25 \\
         &      & Random edges & $\downarrow$ 9.88\% & $\downarrow$ 18.97\% & $\downarrow$ 50.44\% & $\uparrow$ 303301.05 \\
\cline{2-7}
         & 5\%  & Bias edges   & $\downarrow$ 35.48\% & $\downarrow$ 17.24\% & $\downarrow$ 58.41\% & $\uparrow$ 47092.63 \\
         &      & Random edges & $\downarrow$ 10.03\% & $\downarrow$ 18.97\% & $\downarrow$ 54.87\% & $\uparrow$ 77022.32 \\
\cline{2-7}
         & 10\% & Bias edges   & $\downarrow$ 35.43\% & $\downarrow$ 17.24\% & $\downarrow$ 61.06\% & $\uparrow$ 210336.20 \\
         &      & Random edges & $\downarrow$ 9.21\% & $\downarrow$ 18.97\% & $\downarrow$ 55.75\% & $\uparrow$ 340557.07 \\
\hline
\end{tabular}
\caption[.]{
\begin{textblock}
\item
  Change in bias and downstream performance when ablating the top-$K\%$ EAP-scored (bias) edges
  versus an equal-cardinality set of randomly sampled highly-active edges (random control).
\item
  Bias is reported as percentage reduction;
\item
  CoLA and HellaSwag as percentage change;
\item
  WikiText perplexity as absolute increase.
\item
  The evaluation benchmarks correspond to
  linguistic acceptability (CoLA)~\cite{warstadt2018neural},
  commonsense continuation on HellaSwag~\cite{Zellers2019HellaSwagCA},
  and
  language modeling quality (WikiText)~\cite{merity2016pointer}.
\end{textblock}
}
\label{tab:edge_removal_tradeoff}
\end{table*}

\end{document}